\documentclass[preprint, dvipsnames]{elsarticle}
\journal{arXiv}
\usepackage{tikz}
\usepackage{multirow}
\usepackage{threeparttable}
\usepackage{xcolor}
\usepackage{graphics}
\usepackage{booktabs}
\usepackage{hyperref}
\usepackage{xurl}
\usepackage{todonotes}
\usepackage{array}
\usepackage{rotating}
\newcolumntype{L}[1]{>{\raggedright\let\newline\\\arraybackslash\hspace{0pt}}m{#1}}
\newcolumntype{R}[1]{>{\raggedleft\let\newline\\\arraybackslash\hspace{0pt}}m{#1}}
\newcolumntype{C}[1]{>{\centering\let\newline\\\arraybackslash\hspace{0pt}}m{#1}}

\newcommand{\samplesize}{1963}
\newcommand{\samplesizecleaned}{1889}

\newcommand*\circled[1]{\tikz[baseline=(char.base)]{
            \node[shape=circle,fill,inner sep=1pt] (char) {\textcolor{white}{#1}};}}

\AtBeginDocument{%
  \providecommand\BibTeX{{%
    \normalfont B\kern-0.5em{\scshape i\kern-0.25em b}\kern-0.8em\TeX}}}

\newcommand{\tl}[1]{\textcolor{RubineRed}{[TL: #1]}}

\begin{document}

\begin{frontmatter}



\title{What Makes People Install a COVID-19 Contact-Tracing App? Understanding the Influence of App Design and Individual Difference on Contact-Tracing App Adoption Intention}


\author[1]{Tianshi Li \corref{cor1}}
\cortext[cor1]{Corresponding author}
\author[1]{Camille Cobb}
\author[2]{Jackie (Junrui) Yang}
\author[1]{Sagar Baviskar}

\author[1]{Yuvraj Agarwal}
\author[1]{Beibei Li}
\author[1]{Lujo Bauer}
\author[1]{Jason I. Hong}

\affiliation[1]{organization={Carnegie Mellon University},
            addressline={5000 Forbes Avenue},
            city={Pittsburgh},
            postcode={15213},
            state={PA},
            country={United States}}
\affiliation[2]{organization={Stanford University},
            addressline={450 Jane Stanford Way},
            city={Stanford},
            postcode={94305},
            state={CA},
            country={United States}}

\begin{abstract}
Smartphone-based contact-tracing apps are a promising solution to help scale up the conventional contact-tracing process.
However, low adoption rates have become a major issue that prevents these apps from achieving their full potential.
In this paper, we present a national-scale survey experiment ($N = \samplesize$) in the U.S.\ to investigate the effects of app design choices and individual differences on COVID-19 contact-tracing app adoption intentions.
We found that individual differences such as prosocialness, COVID-19 risk perceptions, general privacy concerns, technology readiness, and demographic factors played a more important role than app design choices such as decentralized design vs.\ centralized design, location use, app providers, and the presentation of security risks.
Certain app designs could exacerbate the different preferences in different sub-populations which may lead to an inequality of acceptance to certain app design choices (e.g., developed by state health authorities vs.\ a large tech company) among different groups of people (e.g., people living in rural areas vs.\ people living in urban areas).
Our mediation analysis showed that one's perception of the public health benefits offered by the app and the adoption willingness of other people had a larger effect in explaining the observed effects of app design choices and individual differences than one's perception of the app's security and privacy risks.
With these findings, we discuss practical implications on the design, marketing, and deployment of COVID-19 contact-tracing apps in the U.S.
\end{abstract}



\end{frontmatter}

\section{Introduction}
The COVID-19 pandemic started in March 2020 and has killed 316,844 people in the US as of December 21, 2020\footnote{\url{https://web.archive.org/web/20201222043017/https://covid.cdc.gov/covid-data-tracker/\#cases_casesper100klast7days}}.
Contact tracing is widely known as a key strategy to slow the spread of infectious diseases such as COVID-19.
It involves identifying who may have been exposed to an infected person and helping exposed people take protective measures at the right time~\cite{Prioriti31:online}.
The conventional approach to contact tracing relies on manual investigation, which can not keep up with the rising cases during the global COVID-19 outbreak~\cite{SurveyOf88:online, ContactT19:online}.
Hence, smartphone-based contact-tracing apps have been proposed as a complementary solution to help scale up the contact tracing process~\cite{troncoso2020decentralized, chan2020pact, rivest2020pact}.
The effectiveness of contact-tracing apps is contingent on a critical fraction of the population installing and using the app~\cite{hinch2020effective, ferretti2020quantifying}.
However, deployed contact-tracing apps have suffered from low adoption rates (from 21.6\% in Australia to 0.2\% in Philippines)~\cite{COVID1987:online}, with security and privacy concerns blamed as a main culprit~\cite{braithwaite2020automated}.

Although recent research has investigated factors that affect people's willingness to install a contact-tracing app in general~\cite{abeler2020support, redmilesuser, horstmann2020does, walrave2020ready, walrave2020adoption, o2020national, bachtiger2020belief, abuhammad2020covid, von2020covid, hassandoust2020individuals, altmann2020acceptability, saw2020predicting, simko2020covid, kostka2020times, trang2020one, zhang2020americans}, some important aspects remain unclear. In particular, we aim to focus on three fundamental issues:

First, the design of contact-tracing apps lends itself to multiple choices featuring different trade-offs between security/privacy risks and public health benefits~\cite{baumgartner2020mind, redmilesuser, cho2020contact, ahmed2020survey, martin2020demystifying, shubina2020survey}.
Researchers have conducted choice-based conjoint studies to measure user preferences of different configurations of COVID-19 contact-tracing apps for the UK population  \cite{horvath2020citizens, wiertz2020predicted} and the U.S. population \cite{li2020decentralized, zhang2020americans}.
This method emulates a situation where users have to choose between app designs, but the nature of contact-tracing apps determines that users in a certain region can only choose to install or not install a single designated app\footnote{Currently, every country/region only has one active version of contact-tracing app~\cite{bay2020bluetrace, Howtoget92:online}}.

Second, prior research in contact-tracing apps focuses solely on measuring people's general intentions to \textit{install} the app~\cite{abeler2020support, redmilesuser, horstmann2020does, walrave2020ready, walrave2020adoption, o2020national, bachtiger2020belief, abuhammad2020covid, von2020covid, hassandoust2020individuals, altmann2020acceptability, saw2020predicting, simko2020covid, kostka2020times, trang2020one, zhang2020americans}. However, app installation intentions is not sufficient for effective contact tracing because users must also \textit{actively report cases} and \textit{keep the app installed} in the long run~\cite{IrishBattery:online}.

Third, previous research has conducted qualitative studies to identify reasons why people would or would not install a contact-tracing app~\cite{simko2020covid, li2020decentralized, williams2020public, abeler2020support}, with perceived risks and benefits turning out to be recurring themes.
However, there is a lack of quantitative understanding in how perceived risks and benefits vary with different app designs, across people, and how these variances affect the app adoption intentions accordingly. As a result, it remains unclear which app designs best reconcile the risk-benefit trade-offs and what are the rationales behind the preferences of different sub-populations.

In this paper, we present a national survey experiment ($N=\samplesize{}$) in the U.S.\ to complement prior findings on the impact of app design choices on app adoption intention for contact-tracing apps. We focus primarily on three research questions:
\begin{description}
\item[RQ1] To what extent do app design choices affect people's adoption intentions about a COVID-19 contact-tracing app?
\item[RQ2] To what extent do individual differences affect people's adoption intentions about a COVID-19 contact-tracing app?
\item[RQ3] How do people's perceived risks and benefits about a contact-tracing app mediate the influence of app design choices and individual differences on the app adoption intention?
\end{description}

In our study, we used a between-subjects factorial design, showing each participant only one solution and asking about their intentions to install and use the app.
This is a better approximation for the choice they will actually face and can therefore lead to a more realistic estimation of how app design differences shape adoption intentions, compared to previous studies that have used a within-subjects approach.
We vary design decisions by controlling four variables: proximity-based contact tracing architecture (i.e., decentralized vs. centralized architecture), location use, app provider, and security risk presentation.
The first three correspond to app design choices that were found to be important in prior research in building privacy-preserving contact-tracing apps.
The fourth variable \textit{security risk presentation} allows us to compare participants' adoption intentions when not primed about any security risks and when primed about one of three major security risks of contact-tracing apps: data breach risk, secondary data use risk, and the re-identification risk of COVID-19 positive users.
We also requested participants to answer questions about personal characteristics (prosocialness, COVID-19 risk perceptions, general privacy concerns, and technology readiness) and demographic information (e.g., age, gender) for analyzing the effects of individual differences on adoption intentions.
Our study resulted in a number of key findings, including:

\begin{itemize}
    \item 58.9\% people reported that they at least somewhat agreed to install the app, which is similar to prior work's estimation such as 55\% in~\citet{li2020decentralized} and 59\% in a national poll~\cite{Washingt23:online}. However, only 41.7\% people reported they at least somewhat agreed that most people would install this app, which shows that U.S.\ people hold an overly pessimistic attitude towards the adoption of contact-tracing apps. 76.2\% people reported that they at least somewhat agreed to report to the app if they test positive. This suggests that people are more amenable to using contact-tracing apps and contribute their data when they test positive for COVID-19.
    \item App design choices had very small effects on all five aspects of app adoption intention (e.g., install app, report positive cases, keep the app installed). People were significantly more inclined to install apps that collect location than apps that do not collect location due to the additional benefits from the location data (e.g., for analyzing infection hotspots). Among the three security risks we tested, all of them increased users' perceived security and privacy risks while only the secondary data use risk significantly reduced adoption intention.
    \item Individual differences had large effects on all five aspects of app adoption intention. Older people, females, and essential workers were significantly less inclined to install a COVID-19 contact-tracing app, while Hispanics, people with higher household income, frequent public-transit users during the pandemic, and people living in urban areas were significantly more inclined to install a COVID-19 contact-tracing app.
    \item Certain app design choices could exacerbate the difference in adoption intention due to individual differences, which could lead to potentially unbalanced adoption for certain sub-populations. For example, people living in urban areas showed similar acceptance of state health authorities as the app provider and of a large tech company as the app provider, while people living in rural areas showed much lower acceptance of a large tech company than of state health authorities.
    \item The perceptions about the app's benefits and how much adoption the app can achieve played a more important role in determining one's intention to install a contact-tracing app than perceptions about security and privacy risks.
\end{itemize}

\section{Related Work and Research Questions}

In this section, we present an overview of the contact-tracing app design space that we are studying in the survey experiment, drawing on both research proposals and industry frameworks (e.g., the Google/Apple Exposure Notification API) and review findings of prior work to introduce our research questions.

\subsection{Contact-Tracing App Adoption Challenge}
\label{sec:app_adoption_aspects}
A contact-tracing app needs widespread adoption to work~\cite{hinch2020effective, ferretti2020quantifying}.
Specifically, the installation rate has been widely used as a success metric of contact-tracing apps~\cite{WhyArent70:online} and previous research focused on estimating the percentage of people that will install contact-tracing apps~\cite{redmilesuser, Washingt23:online} and factors that affect people's willingness to install contact-tracing apps~\cite{li2020decentralized, zhang2020americans, horvath2020citizens, wiertz2020predicted, simko2020covid}.
However, for continued contact-tracing users need to keep the app installed and actively report if they test positive~\cite{FrancesC22:online}.
Some evidence has demonstrated that long-term use of the app and honest reporting of positive cases could be impeded by usability concerns (e.g., shorter battery life~\cite{Covid19a10:online, redmilesuser}) and privacy concerns (e.g., the app could remain as a surveillance tool after the pandemic~\cite{zhang2020americans}).
Note that the usability and privacy issues vary greatly among different app designs.

To provide a more comprehensive understanding about factors that affect the adoption intentions of contact-tracing apps, we measure five \textit{outcome variables} of different aspects of adoption in our survey design and analysis, including \circled{1}~the general app installation intentions \circled{2}~whether to report to the app if the user tests positive for COVID-19, \circled{3}~whether to keep the app installed when the battery drains faster, \circled{4}~when COVID-19 cases are steadily decreasing and \circled{5}~when a vaccine becomes available).

\subsection{Effects of App Design Choices on Contact-Tracing App Adoption Intentions}
\label{sec:design_space_overview}

Many digital technologies have been proposed and deployed to help combat the pandemic.
In this paper, we focus on smartphone contact-tracing apps that users voluntarily install to complement conventional contact tracing~\cite{PrivacyP48:online}.
Contact-tracing apps are inherently privacy-sensitive as they rely on users' sensitive data such as their contact history and location history to function~\cite{cho2020contact, baumgartner2020mind}.
On the other hand, collecting more data can improve the accuracy of the automated contact tracing results~\cite{redmilesuser} and provide more information to health workers~\cite{ivers2020can, horvath2020citizens}.
To tackle this risk-benefit trade-off issue, researchers have proposed technical solutions for privacy-preserving contact tracing.
In the following, we introduce two main design dimensions of contact-tracing apps: Proximity-based contact tracing and Location-based contact tracing. Then we discuss two other factors related to contact-tracing app design: app providers and security risks.
Research questions proposed in this subsection are extensions of RQ1: ``\textit{To what extent do app design choices affect people’s adoption intentions about a COVID-19 contact-tracing app?}''


\subsubsection{Proximity-Based Contact Tracing}
Most contact-tracing apps offer Bluetooth Low Energy (BLE)-based proximity tracking to notify people who have recently come into close contact with people who test positive for COVID-19~\cite{redmilesuser, ahmed2020survey, martin2020demystifying, shubina2020survey}.
In March 2020, Singapore created the first COVID-19 contact-tracing app using a \textit{centralized} architecture which completes the contact-tracing process on the server end~\cite{bay2020bluetrace}.
This approach can lead to severe security risks because users' identities (e.g., phone numbers) are associated with their COVID-19 exposure status~\cite{bay2020bluetrace, cho2020contact}.

Therefore, many researchers have proposed \textit{decentralized} architectures that can fulfill the fundamental need of sending exposure notifications to people who might be infected with minimum data shared with a central entity ~\cite{troncoso2020decentralized, chan2020pact, rivest2020pact}.
This allows users to remain anonymous from the central server, but there is still a risk that other app users can identify the infected user they were exposed to by installing modded app that logs additional information such as locations~\cite{ahmed2020survey, troncoso2020decentralized} along with the exposure history.
Because the contact-tracing process is completed on the users' phones, the central server does not know how many exposure notifications were sent to users and how users reacted to them.
This makes it difficult to evaluate the efficacy of the system and integrate it with the conventional contact tracing to facilitate further testing and quarantine processes~\cite{ivers2020can, horvath2020citizens}.
That being said, Google and Apple used this architecture in their Google-Apple Exposure Notification (GAEN) framework~\cite{Exposure87:online, PrivacyP48:online}, which has become the most prevalent way of building contact-tracing apps in the U.S.~\cite{Howtoget92:online}

Researchers have also proposed \textit{privacy-preserving centralized} contact-tracing architectures~\cite{peppptdo31:online, castelluccia2020robert}. Like decentralized architectures, these allow users to remain anonymous from the central server. 
Because the contact-tracing process is completed on the server end, the central server can track when and how many exposure notifications are sent out to help measure the performance of the system and integrate with the conventional contact tracing.
However, it is still possible for app providers to infer the identities of users using the anonymized contact history shared with the server~\cite{troncoso2020decentralized}.
This system could also suffer from the re-identification risk under a Sybil attack, namely users of this app can narrow down the scope of infected users they were exposed to by registering multiple accounts~\cite{troncoso2020decentralized}.

\citet{li2020decentralized} conducted a choice-based conjoint study that studied similar design choices and found users preferred the centralized architecture.
However they did not investigate the \textit{privacy-preserving centralized} architecture which serves as a middle ground between the two extreme solutions. Besides, their description highlighted the re-identification risk for the decentralized architecture but did not mention other risks like data breach that a centralized architecture is more susceptible to, which could bias users' decisions.
In our study, we examine users' preferences and feelings about these three mechanisms of proximity-based contact tracing described above. 

\begin{description}
\item[RQ1.1] To what extent do different proximity-based contact tracing designs (1. decentralized architecture, 2. centralized architecture using anonymized identifiers, 3. centralized architecture using real identities) affect people's intentions to adopt a COVID-19 contact-tracing app?
\end{description}

\subsubsection{Location Use}

Infected people's location histories are useful for contact-tracing, especially for tracing indirect contact (e.g., spread through shared surfaces or aerosol in a public spaces) which cannot be captured by proximity-based contact-tracing apps~\cite{culler2020covista, raskar2020apps}.
However, the use of location data in contact-tracing apps has been controversial and the Google/Apple exposure notification framework even forbids apps that built with it to collect location data~\cite{Exposure87:online, PrivacyP48:online} due to the risks of increased surveillance of all app users and privacy leak and stigmatization of infected users~\cite{SouthKor57:online, zhang2020americans}.

Previous research has not reached a consensus on how location use affects users' preferences of contact-tracing apps.
\citet{zhang2020americans} showed that using Bluetooth data for proximity-only contact tracing increases users' acceptance of contact-tracing apps compared to using GPS for location-based tracing, while \citet{li2020decentralized} showed that collecting location data in public areas and providing users with infection hotspot information significantly increased willingness to  adopt.
These findings suggest that location data collection may be more acceptable to users when it provides additional benefits over basic proximity-based contact tracing.
Therefore, our study focuses on comparing no location collection (and no additional benefits) with location features that have additional benefits and can still preserve privacy to some extent.

The first feature we study relies on \textit{storing the location data on device} to mitigate privacy risks, such as the Care19 Diary app in South Dakota, USA~\cite{COVIDinS96:online}.
If a user of the app tests positive for COVID-19, they can refer to the location logs tracked by this app to help them recall their recent whereabouts when interviewed by a human contact tracer.

The second feature we study relies on \textit{uploading the location data of infected users} so that infection hotspots recently visited by many infected users can be shared with the public.
Research has shown that users find knowing about infection hotspots useful and may be more willing to install an app that offers this feature~\cite{redmilesuser, li2020decentralized}.
To protect users' privacy, researchers have proposed technologies such as Safe Paths~\cite{raskar2020apps} that enable users to upload anonymized, redacted, and obfuscated location history.

\begin{description}
\item[RQ1.2] To what extent do different location-based contact tracing features (1. no location use, 2. storing location on device as a memory aide, 3. sharing locations with health authorities to analyze infection hotspots) affect adoption intention for a COVID-19 contact-tracing app?
\end{description}

\subsubsection{App Providers}

In addition to different app designs, the organizations that develop and release the contact-tracing app and therefore have access to users' data can also have significant impact on users' intentions to adopt it ~\cite{redmilesuser, li2020decentralized, zhang2020americans, horvath2020citizens, wiertz2020predicted, simko2020covid}.
Previous research found that sharing sensitive information such as location and contact history with government agencies in general could lead to a low acceptance of contact-tracing apps~\cite{simko2020covid, horvath2020citizens, wiertz2020predicted}.
In contrast, sharing data with health authorities in particular such as CDC of the U.S. and NHS of the U.K. could improve users' willingness to adopt contact-tracing apps~\cite{li2020decentralized, horvath2020citizens, wiertz2020predicted}.

However, the health-authority-leading solution encountered more challenges in the U.S. than other places.
In the US, there is no single national contract tracing app due to the lack of coordination by the federal government, while the rollout of state-specific apps has been slow due to the lack of technical expertise in state health departments~\cite{WhyArent70:online}. In fact, scholars recommended to seek ``the piecemeal creation of public trust'', and other entities have taken actions to help build contact-tracing apps~\cite{blasimme2020s}.
For example, Google and Apple launched the ``Exposure Notifications Express'' project which integrates contact tracing as an opt-in feature built into their OS's to alleviate the need of users to install any contact-tracing apps~\cite{AppleGoo94:online}.
Similarly, some U.S. universities have built their own contact-tracing apps to protect their faculty, staff, and students on campus~\cite{CMUCreat95:online, UCCampus71:online, UCBerkel10:online}.

In our study, we examine the impact of the four providers mentioned above on the adoption intention: state-level health authorities, federal-level health authorities, a large tech company (such as Google and Apple), and the users' employer or school.

\begin{description}
\item[RQ1.3] To what extent do different app providers (1. state health authorities, 2. federal health authorities, 3. a large tech company, 4. your employer or school) affect people's adoption intentions of a COVID-19 contact-tracing app?
\end{description}

\subsubsection{Security Risks}

Despite all the technical approaches to protect users' privacy, the nature of contact-tracing apps means that some security risks are inevitable regardless of the specific app design, though developers rarely mention them in their app description\cite{baumgartner2020mind, cho2020contact}.
However, very few contact-tracing app studies explicitly explained their security risks to their participants, and they focused on a certain type of security risk that is less protected against in a certain app design.
For example, ~\citet{li2020decentralized} highlighted the re-identification risk of infected users, which decentralized apps are more vulnerable to, and learned users tended to prefer centralized apps over decentralized ones.
~\citet{horvath2020citizens} controlled for whether to prompt users about the data breach risk, which centralized apps are more vulnerable to, and found the data breach stimuli did not change users' preferences for data storage.

In our research, we want to know how users' awareness of security risks affects their decisions in adopting contact-tracing apps.
Because different app design choices are more vulnerable to different risks, we are also interested in whether they have different levels of impact on adoption intention.
Specifically, we test four conditions, including a baseline condition that does not directly mention any security risk, and three other conditions that prime users about the data breach risk, secondary data use risk, or the re-identification risk.

\begin{description}
\item[RQ1.4] To what extent does priming users about different security risks of a COVID-19 contact-tracing app (1. not priming users about security risks, 2. priming about data breach risks, 3. priming about secondary data use risks, 4. priming about re-identification risks) affect their adoption intentions?
\end{description}

\subsection{Effects of Individual Differences on Contact-Tracing App Adoption Intention}
\label{sec:individual_difference_rqs}

Previous research has demonstrated that individual differences can play an important role in people's willingness to adopt a COVID-19 contact-tracing app.
In our survey, we build upon prior findings to examine how different sub-populations and people who hold different opinions to certain topics in general (e.g., privacy, COVID-19 risks, etc.) react to contact-tracing apps.
Research questions proposed in this subsection are extensions of RQ2: ``\textit{To what extent do individual differences affect people’s adoption intentions about a COVID-19 contact-tracing app?}''

\subsubsection{Prosocialness}
Altruism and contributing to the ``greater good'' were identified as important reasons for contact-tracing app supporters~\cite{simko2020covid, williams2020public, redmilesuser}.
Furthermore, \citet{trang2020one} found that emphasizing the societal benefits of the app led to a higher adoption willingness than emphasizing the benefits to users themselves.
Because people who are more prosocial may feel more strongly about contributing to the ``greater good,'', marketing these apps to appeal to this aspect of people could foster adoption and increase overall rates of usage of contact-tracing apps.
Hence we have the following research question to formally study the effects of prosocialness on adoption intentions:

\begin{description}
\item[RQ2.1] To what extent is one's prosocialness associated with COVID-19 contact-tracing app adoption intentions?
\end{description}


\subsubsection{General Privacy Concerns}
In contrast, the fear of increased surveillance and privacy risks were identified as important reasons for people who did not want to install contact-tracing apps~\cite{simko2020covid, williams2020public, redmilesuser, hassandoust2020individuals}.
As people's perceived privacy risks about contact-tracing apps in particular is likely to be affected by their privacy concerns in general, we have the following research question:

\begin{description}
\item[RQ2.2] To what extent is one's general privacy concern associated with COVID-19 contact-tracing app adoption intentions?
\end{description}

\subsubsection{COVID-19 Risk Perception}
We learned from past pandemics that public perceptions of the risks of a disease has a significant influence on the success of controlling the spread of a highly infectious disease~\cite{dryhurst2020risk, epstein2008coupled}.
However, conspiracy theories about the seriousness of COVID-19 have become barriers to the adoption of measures to control the spread of the disease such as social distancing~\cite{romer2020conspiracy}.
As a result, we have the following research question.

\begin{description}
\item[RQ2.3] To what extent is one's risk perception about COVID-19 associated with COVID-19 contact-tracing app adoption intentions?
\end{description}

\subsubsection{Technology Readiness}
\citet{parasuraman2015updated} divided people into five segments based on their attitudes towards technologies, including \textit{Skeptics}, \textit{Explorers}, \textit{Avoiders}, \textit{Pioneers}, and \textit{Hesitators} and found that they exhibit different intentions and behaviors in adopting new technologies.
Because contact-tracing apps are a new technology designed to complement the conventional manual contact-tracing process, people's intrinsic attitudes towards new technologies could have an essential impact on their adoption of contact-tracing apps.
Therefore, we have the following research question:

\begin{description}
\item[RQ2.4] To what extent is one's attitude towards new technologies associated with COVID-19 contact-tracing app adoption intentions?
\end{description}

\subsubsection{Demographics}
A large body of research has studied the influence of demographic factors such as age~\cite{horstmann2020does, walrave2020ready, hassandoust2020individuals, walrave2020adoption}, gender~\cite{horstmann2020does, walrave2020ready, hassandoust2020individuals, walrave2020adoption}, race~\cite{anderson2020most}, education~\cite{horstmann2020does, walrave2020ready, hassandoust2020individuals, walrave2020adoption}, income~\cite{abuhammad2020covid}, and living area~\cite{abuhammad2020covid} on COVID-19 contact-tracing app adoption intentions in the settings of various countries.
However, their findings are not consistent.
For example, regarding the age factor, some research showed that older people are significantly less willing to adopt contact-tracing apps~\cite{horstmann2020does, von2020covid}, while some found an opposite trend~\cite{hassandoust2020individuals} and some did not find that age had a significant influence~\cite{walrave2020ready, walrave2020adoption, saw2020predicting}.
The difference could be due to differences in culture, political climate, and the stage of the pandemic in different countries when the studies were conducted.
It could also be related to the difference in study design (e.g., within-subjects vs. between-subjects design) and the app description (e.g., a general description vs. a detailed description of the risks and benefits of a specific design).

\begin{description}
\item[RQ2.5] To what extent do demographic factors (e.g., age, gender, race, education, income, living area) correlate with a person's willingness to adopt a COVID-19 contact-tracing app in the U.S.?
\end{description}

Note that certain sub-populations are at higher risks to get exposed to COVID-19, such as essential workers, health workers, and people who need to take public transit frequently during the pandemic.
However, there has been little research about the adoption of contact-tracing apps for these people.
Therefore, our survey asks users to self-report whether they belong to any of the above high-risk sub-populations to answer the following research question:

\begin{description}
\item[RQ2.6] To what extent do people at higher risks of getting exposed to COVID-19 (e.g., essential workers, health workers, frequent public transit users) like to install a COVID-19 contact-tracing app?
\end{description}

Although some past work examined people's reactions to different app designs~\cite{horvath2020citizens, wiertz2020predicted, li2020decentralized, zhang2020americans, utz2020apps}, they focused on finding the designs that are likely to achieve high adoption rate for the entire population.
We want to take a step further to understand more nuances about how installation intentions of different sub-populations (e.g. men vs. women, older people vs. younger people) are moderated by different app design choices.
Hence, the following research question studies the interaction effect between factors related to app design choices and demographic factors:

\begin{description}
\item[RQ2.7] To what extent do app design choices moderate the intentions to install a COVID-19 contact-tracing app of different sub-populations?
\end{description}

\subsection{Explaining the Effects of App Design Choices and Individual Differences on Installation Intentions Through Risk-Benefit Tradeoffs}
\label{sec:mediation_rqs}
Recent qualitative research has identified the \textit{risks} of increased surveillance and privacy invasion and the \textit{benefits} to society and to the users themselves as two main reasons that explain why a person would install or not install a COVID-19 contact-tracing app~\cite{simko2020covid, li2020decentralized, williams2020public, abeler2020support}.
These findings are in line with the Privacy Calculus theory~\cite{dinev2006extended}, which states that individuals view privacy as a trade-off problem and make data disclosure decisions by weighing the potential risks and potential benefits.
Correspondingly, some prior work has drawn on the Privacy Calculus theory and examined the influence of perceived risks and benefits in users' decisions and how perceived risks and benefits mediate the relationship between abstract attributes and app adoption intentions.
Specifically, \citet{hassandoust2020individuals} drew on the Privacy Calculus theory and conducted structural equation modeling to examine the influence of perceived risks and benefits in users' decisions and found that technical attributes (\textit{anonymity} and \textit{information sensitivity}) could influence the adoption intentions by affecting users' risk beliefs.
Despite the theoretical insights, it is hard to link these abstract features to existing app designs and translate the results to practical design recommendations.

In our survey, we use a similar method as the above work~\cite{hassandoust2020individuals} to further explain why certain app design choices and individual differences have significant influences on app installation intention.
We also use perceived risks and benefits as mediators, while our independent variables include factors related to app design choices grounded in real-world contact-tracing app designs (Section~\ref{sec:design_space_overview}) rather than abstract features, which can more directly contribute to our understanding of the design space. The following research questions are extensions of RQ3: ``\textit{How do people's perceived risks and benefits about a contact-tracing app mediate the influence of app design choices and individual differences on the app adoption intention?}'':

\begin{description}
\item[RQ3.1] (\textit{Risks}) To what extent do security and privacy risks mediate the relationship between independent variables (i.e., app design choices and individual differences) and the installation intention of a COVID-19 contact-tracing app?
\item[RQ3.2] (\textit{Self benefits}) To what extent does perceived protection to the users themselves mediate the relationship between independent variables (i.e., app design choices and individual differences) and the installation intention of a COVID-19 contact-tracing app?
\item[RQ3.3] (\textit{Societal benefits}) To what extent does perceived effectiveness in slowing the spread of COVID-19 mediate the relationship between independent variables (i.e., app design choices and individual differences) and the installation intention of a COVID-19 contact-tracing app?
\end{description}

Due to the unique requirement of achieving a widespread adoption to be effective, how much a person believes other people would like to install the app could affect their perception of the efficacy of the app~\cite{li2020decentralized, utz2020apps}.
Therefore, we also include the factor \textit{perceived adoption} as a potential mediator in our analysis:

\begin{description}
\item[RQ3.4] (\textit{Perceived adoption}) To what extent does perceived adoption of the app mediate the relationship between independent variables (i.e., app design choices and individual differences) and the installation intention of a COVID-19 contact-tracing app?
\end{description}

\section{Methodology}
To answer the research questions and test the hypotheses about factors that affect people's intentions to adopt a COVID-19 contact-tracing app, we conducted a randomized between-subjects survey experiment on a representative sample of U.S. population ($N=\samplesize{}$) recruited using a Qualtrics panel.
The sample size was determined before the formal study based on power analysis results ($\beta > 0.8$).
The effect size was estimated using data collected in pilot studies.
Our survey was programmed and hosted on Qualtrics.
The data was collected in November, 2020.
Our study has been reviewed and approved by our institution's IRB.

\subsection{Participants}
We recruited participants based in the U.S. using a Qualtrics online panel.
To obtain a nationally representative sample, we employed a quota-sampling method~\cite{cumming1990probability} for recruiting participants and controlled for gender, age, race, and living region to make the distributions of these variables consistent with U.S. census data.
We required participants to be fluent English speakers, aged 18 or older, and use smartphones.
Qualtrics handled the entire data collection process, including recruiting, survey distribution, and compensation. We paid \$6.5 for each complete response.

We obtained 2026 responses that passed all understanding check and attention check questions using a Qualtrics online panel\footnote{\url{https://web.archive.org/web/20201120174828/https://www.qualtrics.com/research-services/online-sample/}}.
63 responses were removed as they did not provide a valid ZIP code, which yields a final sample of \samplesize{} unique responses.
The survey was configured to allow a respondent to take the survey only once so they could not re-attempt the survey after failing attention checks.

\subsection{Experiment Design}

\begin{table}[htbp]
    \centering
    \caption{Summary of the experimental manipulations to participants}
    \resizebox{\linewidth}{!}{%
    \begin{tabular}{p{0.22\linewidth}p{0.34\linewidth}p{0.5\linewidth}}
    \toprule
    Manipulations    & Conditions & App behaviors and data practices\\
    \hline
    \multirow{3}{3cm}{Proximity-based contact tracing (RQ1.1)} & Decentralized & Notify exposed users.\newline Contact tracing on device using anonymous IDs. \newline\\
         & Anonymized Centralized & Notify exposed users. \newline
        Provide health authorities with exposure stats. \newline
        Contact tracing on servers using anonymous IDs. \newline\\
         & Identified Centralized& Notify exposed users. \newline
        Provide health authorities with exposure stats. \newline
        Support health workers to contact exposed users. \newline
        Contact tracing on servers using real identities.\\
    \hline
    \multirow{3}{2.8cm}{Location use (RQ1.2)}     & No location use & No location history will be collected. \newline\\
         & Location on device & Help infected users recall their recent whereabouts. \newline
         Location history stored on device. \newline\\
         & Location uploaded & Help infected users recall their recent whereabouts. \newline
         Help health workers analyze hotspots of infection. \newline
         Infected users' location history stored on servers.\\
    \hline
    \multirow{3}{2.8cm}{App provider (RQ1.3)}     & State health authorities & State health authorities built the app. \\
         & Federal health authorities & Federal health authorities built the app. \\
         & Tech company & A large tech company built the app. \\
         & Employer or school & Your employer or school built the app. \\
    \hline
    \multirow{3}{2.8cm}{Security risk (RQ1.4)}     & No security risk & No security risk is mentioned.\\
         & Data breach risk & Stored data may be or stolen by outside hackers.\\
         & Secondary use risk & Data may be stored longer than needed and used for other purposes.\\
         & Re-identification risk & Exposed users could guess who were infected and led to their exposure.\\
    \bottomrule
    \end{tabular}}
    \label{tab:experimental_design}
\end{table}

As summarized in Table \ref{tab:experimental_design}, our study follows a 3 (Decentralized vs. Anonymized Centralized vs. Identified Centralized) x 3 (No location use vs. Location on device vs. Location uploaded) x 4 (State health authorities vs. Federal health authorities vs. Tech company vs. Employer or school) x 4 (No security risk vs. Data breach risk vs. Secondary data use risk vs. Re-identification risk) factorial design.
Each participant was randomly assigned into one condition and saw the app description created with the selected values of the four variables.
Note that compared to prior work, in this study, each participant is only presented with one contact-tracing app design to simulate a more realistic setting and reduce the effect of fatigue.
Then they reported their willingness to install and use the app and their perceived risks and benefits of the app.
These manipulations allow us to study the effects of the four factors related to app design choices on the adoption intentions for contact-tracing apps (RQ1.1-1.4, see Section~\ref{sec:design_space_overview}) and study the how app design choices affect the adoption intentions through perceived risks and benefits (RQ3.1-3.4, see Section~\ref{sec:mediation_rqs}).
We intentionally had each participant see only one app design to emulate the real-world situation when there is only one COVID-19 contact-tracing app available in a region.
This design also reduces the potential fatigue caused by reading and evaluating multiple app designs.

We also asked participants to provide their demographic information, which allows us to study the effects of individual differences on the adoption intentions for contact-tracing apps (RQ2.1-2.4, RQ2.5 and RQ2.6, see Section~\ref{sec:individual_difference_rqs}) and the interaction effects between app design choices and individual differences (RQ2.7, see Section~\ref{sec:individual_difference_rqs}).

\subsection{Experiment Procedure}

\begin{figure}
    \centering
    \includegraphics[width=1\linewidth]{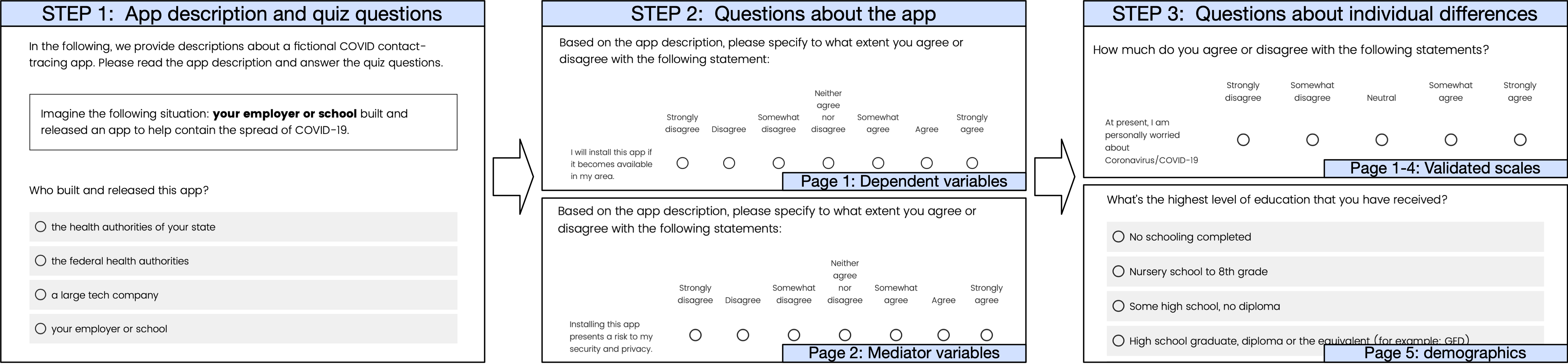}
    \caption{An Illustration of the experiment procedure. Our experiment consists of three main steps. The first step presents the app description and requires participants to correctly answer all quiz questions to proceed. The second step asks participants to report their intentions to install and use the app and their perceived risks, benefits, and community adoption rate of this app. The third step asks questions about the participants themselves, including validated scales that measure personal characteristics such as prosocialness and common demographic questions.}
    \label{fig:experiment_procedure}
\end{figure}

Our experiment consisted of three steps as demonstrated in Figure~\ref{fig:experiment_procedure}.
An example of the complete survey can be found at \url{https://github.com/covid19-hcct/HCCT-documents/blob/master/national_survey_design_example.pdf}.

\subsubsection{Step 1: App Description and Quiz Questions} Participants were first presented with a description about the COVID-19 contact-tracing app randomly selected from 144 variations (3x3x4x4 factorial design).
We include a screenshot of one of the app descriptions as a example in the appendices (Figure~\ref{fig:example_app_description}).
To ensure participants correctly understood the app's features and data practices, we required participants to answer quiz questions. If the participants gave an incorrect answer, they could go back to read the description again. However, they could not proceed to the next step until they answered all the quiz questions correctly.
This method is borrowed from previous research that had similar experiment design~\cite{wang2020factors}.
All quiz questions are multiple choice questions except for the questions about security risks which requested participants to type the name of the security risk (ignoring spaces and case differences).
This is because we did not want to prime users in the ``\textit{No security risk}'' condition (control condition) about any security risk from reading the options in the quiz question.

\subsubsection{Step 2: Questions About the App}
\label{sec:step2}
This step contains two pages and both pages began with the same app description as Section 1.
In the first page, participants were asked to answer questions about their intentions to install and use the app.
There were five questions corresponding the five aspects of app adoption introduced in Section~\ref{sec:app_adoption_aspects}, which covered the general intentions to install the app, and intentions to report positive case to the app and keep the app installed.
In the second page, participants were asked to rate their perceived risks, benefits, and other people's adoption intentions.
We inserted an attention check question after all other questions (``This is an attention check, please go ahead and select strongly agree'').
When clicking on the next page button, the survey would automatically terminate if the participants did not pass the attention check.
At the end of this step, there was an open-ended question that allowed participants to freely express their opinions regarding the contact-tracing app.

\subsubsection{Step 3: Questions About Individual Differences}
After answering app-related questions, participants were asked to fill out validated scales that measure their prosocialness~\cite{caprara2005new}, general privacy concerns~\cite{malhotra2004internet}, technology readiness~\cite{parasuraman2015updated}, and COVID-19 risk perceptions~\cite{dryhurst2020risk}.
The four scales were presented in four different pages in random order.
We inserted an attention check question similar to Step 2 for each scale and the survey would terminate when participants clicking the next page button if they failed the attention check on that page.
Finally, participants were asked to fill out demographic questions (e.g., age, gender, race).
The complete list of demographic factors can be found in Section~\ref{sec:demographics_operationalization}.

\subsection{Operationalization}

\subsubsection{Dependent Variables}
\label{sec:dependent_vars}
We asked participants to report their adoption intentions in five aspects on a 7-point likert scale (1=strongly disagree, 7=strongly agree) in Step 2 Page 1 (Section~\ref{sec:step2}).

\textbf{Install app:} We asked participants to rate to what extent they agreed or disagreed with the statement ``I will install this app if it becomes available in my area.''

Then we asked participants to assume they have already installed the app, and then rate to what extent to agreed or disagreed with the following statements:

\textbf{Report positive case}: ``I will report to this app if I test positive.''

\textbf{Shorter battery life}: ``I will keep this app installed even if my phone battery seems to last less long.''

\textbf{Fewer cases}: ``I will keep this app installed even if COVID-19 cases are steadily decreasing in my area.''

\textbf{Vaccine available}: ``I will keep this app installed even if a COVID-19 vaccine becomes widely available.''

\subsubsection{Mediator Variables}
\label{sec:mediator_vars}

We asked participants their perceived risks, benefits, and other people's adoption intentions about the contact-tracing app presented to them on a 7-point likert scale (1=strongly disagree, 7=strongly agree) in Step 2 Page 2 (Section~\ref{sec:step2}).
The statements for each variable are listed as follows:

\textbf{Security and privacy risks}: ``Installing this app presents a risk to my security and privacy.''

\textbf{Self benefits}: ``Installing this app helps me protect myself against COVID-19.''

\textbf{Societal benefits}: ``This app helps slow the spread of COVID-19 in my area.''

\textbf{Perceived adoption}: ``Most people in my area would install this app if it became available.''

\subsubsection{Independent Variables}
\label{sec:independent_vars}
For factors related to app design choices, each presented contact-tracing app description was coded using four variables.
We chose the condition ``\textit{Decentralized}, \textit{No location use}, \textit{State health authorities} developed, \textit{No security risk} mentioned'' as the reference levels for the four variables.
Because they correspond with how contact-tracing apps are built in the U.S. (until December 2020): Different apps are developed for each state using the Google/Apple Exposure Notification framework, which implements the decentralized architecture and forbid the use of location in the same app.

\textbf{Proximity-based contact tracing}: We operationalize the three types of designs as two indicator variables: \textit{Anonymized Centralized} and \textit{Identified Centralized}, which take the value of 1 for participants in the respective condition and 0 otherwise.

\textbf{Location use}: We operationalize the three types of designs as two indicator variables: \textit{Location on device} and \textit{Location uploaded}, which take the value of 1 for participants in the respective condition and 0 otherwise.

\textbf{App providers}: We operationalize the four app provider options as three indicator variables: \textit{Federal health authorities}, \textit{Tech company}, and \textit{Employer or school}, which take the value of 1 for participants in the respective condition and 0 otherwise.

\textbf{Security risks}: We operationalize the four types of designs as three indicator variables: \textit{Data breach risk}, \textit{Secondary use risk}, and \textit{Re-identification risk} which take the value of 1 for participants in the respective condition and 0 otherwise.

For individual differences, we first used validated scales to measure the following personal characteristics of interest:

\textbf{Prosocialness}: We used the 16-item scale developed by \citet{caprara2005new} to measure participants' prosocialness.
The sixteen questions are on a 5-point likert scale and higher score means higher prosocialness.
We define the prosocialness value for each individual as the average rating of the 16 questions so the range of this variable is still $[1, 5]$.
The internal consistency (Cronbach’s alpha) of all 16 questions was 0.94 on our sample which showed high reliability.

\textbf{General Privacy Concerns}: We used the 10-item Internet Users' Information Privacy Concerns (IUIPC) scale developed by \citet{malhotra2004internet} to measure participants' general privacy concerns.
The ten questions are on a 7-point likert scale and higher score means higher privacy concerns.
We define the general privacy concern value for each individual as the average rating of the 10 questions so the range of this variable is still $[1, 7]$.
The internal consistency (Cronbach’s alpha) of all 10 questions was 0.86 on our sample which showed high reliability.

\textbf{COVID-19 Risk Perception}: We developed six questions to measure participants' perceptions about the severity and risks of COVID-19. The questions are adapted from based on \citet{dryhurst2020risk}'s work about COVID-19 risk perceptions.
The six questions are on a 5-point likert scale and higher score means higher COVID-19 Risk Perceptions.
We define the COVID-19 risk perception value for each individual as the average rating of the 6 questions so the range of this variable is still $[1, 5]$.
The internal consistency (Cronbach’s alpha) of all 10 questions was 0.83 on our sample which showed high reliability.

\textbf{Technology readiness}: We used the 16-item Technology Readiness Index (TRI) 2.0 scales developed by \citet{parasuraman2015updated} to measure participants' predisposition to use new technologies.
The sixteen questions are on a 5-point likert scale and high scores indicate positive attitudes to new technologies.
We define the technology readiness value for each individual as the average rating of the 16 questions so the range of this variable is still $[1, 5]$.
The internal consistency (Cronbach’s alpha) of all 10 questions was 0.83 on our sample which showed high reliability.

We also asked users to report demographic factors:
\label{sec:demographics_operationalization}

\textbf{Age}: We provided a text input box to allow participants to enter their age.

\textbf{Gender}: We provided five options for participants to select: ``Male'', ``Female'', ``Non-binary'', ``Prefer not to disclose'', and ``Prefer to self-describe''.
In our regression and mediation analysis, we only included participants who identified themselves as ``Male'' or ``Female'' and code the variable as 1 for ``Female'' and 0 for ``Male'' because the other groups contained too few responses.

\textbf{Race}: We provided 9 options for participants to select: ``American Indian or Alaska Native'', ``Asian'', ``Black or African American'', ``Hispanic or Latino'', ``Middle Eastern'', ``Native Hawaiian or Pacific Islander'', ``White'', ``Prefer not to disclose'', and ``Prefer to self-describe''.
In our regression and mediation analysis, we only included participants who identified themselves as \textit{Asian} and \textit{Black or African American}, \textit{Hispanic or Latino} because the other groups contained too few responses.
We operationalize this variable using three indicator variables: \textit{Asian} and \textit{Black or African American}, \textit{Hispanic or Latino}, which take the value of 1 for participants belonging to the corresponding race and 0 otherwise.

\textbf{Education}: We provided 11 options for participants to select: ``No schooling completed'', ``Nursery school to 8th grade'', ``Some high school, no diploma'', ``High school graduate, diploma or the equivalent (for example: GED)'', ``Some college credit, no degree'', ``Trade/technical/vocational training'', ``Associate degree'', ``Bachelor’s degree'', ``Master’s degree'', ``Professional degree'', ``Doctorate degree''.
Because ``Education'' is an ordinal variable, we converted the 11 options to integers 1 to 11, with 1 corresponding to ``No schooling completed'' and 11 to ``Doctorate degree''.

\textbf{Income}: We provided 7 options for participants to select: ``Less than \$25,000'', ``\$25,000 to \$34,999'', ``\$35,000 to \$49,999'', ``\$50,000 to \$74,999'', ``\$75,000 to \$99,999'', ``\$100,000 to \$149,999'', ``\$150,000 or more''.
Because ``Income'' is an ordinal variable, we converted the 7 options to integers 1 to 7, with 1 corresponding to ``Less than \$25,000'' and 7 to ``\$150,000 or more''.

\textbf{Health workers}: We asked participants to self-report whether they were health workers.
This variable takes the value of 1 for participants who answered ``Yes'', and 0 for “No”.

\textbf{Essential workers} We asked participants to self-report whether they were essential workers\footnote{We provided a definition of essential worker next to the question: ``workers who conduct operations and services that are essential for critical infrastructure operations, such as health care, food service, and public transportation.''}.
This variable takes the value of 1 for participants who answered ``Yes'', and 0 for “No”.

\textbf{Transit use}: We asked the question ``
How often do you take public transportation \textit{during the pandemic}?'' and provided 5 options: ``Never'', ``Rarely'', ``Monthly'', ``More than once a week'', ``Every day''.
Because ``Transit use'' is an ordinal variable, we converted the 5 options to integers 1 to 5, with 1 corresponding to ``Never'' and 5 to ``Every day''.

\textbf{Urban area percentage}: We asked participants to provide their ZIP code to identify which county they resided in when taking the survey.
Then we used the most recent U.S. Census data (2010)\footnote{\url{https://web.archive.org/web/20201210153214/https://www.census.gov/programs-surveys/geography/guidance/geo-areas/urban-rural/2010-urban-rural.html}} to look up what percentage of area of the county is urbanized and operationalize this variable using this number.


\subsection{Statistical Analysis Method}

To answer our research questions, we used two statistical analysis methods: linear regression analysis and mediation analysis.

\subsubsection{RQ1\&2: Linear Regression Analysis}
We created five additive linear regression models to study the main effects of app design choices (RQ1.1-1.4) and individual differences (RQ2.1-2.4, RQ2.5-2.2) for each outcome variable and an interactive linear regression model to study the interaction effects between demographic factors and app design choices (RQ2.7) on the installation intentions. 
Multicollinearity was not a problem for all our linear regression analyese because the maximum generalized variance inflation factors ($GVIF^{(1/(2*Df)})$) for our models is 1.21, which is lower than the cutoff value 2.25.

\subsubsection{RQ3: Mediation Analysis Using Structural Equation Modeling}

To answer RQ3 (Section~\ref{sec:mediation_rqs}), we analyzed the mediation effects of the four mediator variables (Section~\ref{sec:mediator_vars}) using structural equation modeling (SEM), following guidelines from prior literature~\cite{rucker2011mediation, preacher2011effect}.

For our mediation analysis, we focus on the main outcome variable ``Install app'' intention rating.
We first selected independent variables that had a significant effect in our additive linear regression model for this outcome variable.
Then we operationalize our mediation analysis using the following regressions:

\resizebox{\linewidth}{!}{
\begin{tabular}{p{0.01\linewidth}p{1.1\linewidth}}
\\
     1. & Installation intention rating $\sim$ [Selected independent variables] + Security and privacy risk rating + Self benefit rating + Societal benefit rating + Perceived adoption rating\\
     2. & Security and privacy risk rating $\sim$ [Selected independent variables] \\
     3. & Self benefit rating $\sim$ [Selected independent variables] \\
     4. & Societal benefit rating $\sim$ [Selected independent variables] \\
     5. & Perceived adoption rating $\sim$ [Selected independent variables] \\
\end{tabular}}

\section{Results} 

\subsection{Descriptive Statistics}

We summarize the demographics of our survey sample ($N=\samplesize$) in Table~\ref{tab:sample_overview}.
Our sample has consistent demographics statistics with the latest U.S.\ Census data\footnote{
\label{note_us_census}. For age and race, we used \url{https://web.archive.org/web/20201220221336/https://www.census.gov/data/tables/time-series/demo/popest/2010s-national-detail.html}.  
For education, we used \url{https://web.archive.org/web/20201117011544/https://www.census.gov/content/census/en/data/tables/2019/demo/educational-attainment/cps-detailed-tables.html}.
For income, we used \url{https://web.archive.org/web/20201215160528/https://www.census.gov/data/tables/2020/demo/income-poverty/p60-270.html}.}.

\begin{table}[htbp]
    \centering
    \caption{Demographics statistics of our survey sample ($N=\samplesize$). Our sample is consistent with the latest U.S. Census results.}
    \vspace{0.5em}
    \resizebox{\linewidth}{!}{%
    \begin{threeparttable}
    \begin{tabular}{p{0.49\linewidth}R{0.1\linewidth}R{0.2\linewidth}R{0.2\linewidth}}
    \toprule
    Demographic Characteristics & N & Sample (\%) & U.S. (\%) \\
    \midrule
    \textbf{Gender}\\
    \hspace{3mm} Female & 994 & 50.6\% & 51.3\%\\
    \hspace{3mm} Male & 961 & 49.0\% & 48.7\%\\
    \hspace{3mm} Non-binary & 6 & 0.3\% \\
    \hspace{3mm} Prefer not to disclose & 1 & $<$0.1\%\\
    \hspace{3mm} Prefer to self-describe & 1 & $<$0.1\%\\
    \textbf{Age}\\
    \hspace{3mm}18--24 & 171 & 8.7\% & 11.7\%\\
    \hspace{3mm}25--34 & 473 & 24.1\% & 17.9\%\\
    \hspace{3mm}35--44 & 387 & 19.7\% & 16.4\%\\
    \hspace{3mm}45--54 & 245 & 12.5\% & 15.6\%\\
    \hspace{3mm}55--64 & 285 & 14.5\% & 16.4\%\\
    \hspace{3mm}65+ & 402 & 20.5\% & 22.0\%\\
    \textbf{Race}\\
    \hspace{3mm}American Indian or Alaska Native & 20 & 1.0\% & 1.2\%\\
    \hspace{3mm}Asian & 127 & 6.5\% & 6.3\%\\
    \hspace{3mm}Black or African American & 235 & 12.0\% & 13.0\%\\
    \hspace{3mm}Hispanic or Latino & 243 & 12.4\% & 16.8\%\\
    \hspace{3mm}Middle Eastern & 5 & 0.3\%\\
    \hspace{3mm}Native Hawaiian or Pacific Islander & 4 & 0.2\% & 0.2\%\\
    \hspace{3mm}White & 1289 & 65.7\% & 60.1\%\\
    \hspace{3mm}Prefer not to disclose & 11 & 0.5\%\\
    \hspace{3mm}Prefer to self-describe & 29 & 1.5\%\\
    \textbf{Education}\\
    \hspace{3mm} Bachelor's degree or higher & 883 & 45.0\% & 33.3\%\\
    \textbf{Household Income}\\
    \hspace{3mm}Less than \$25,000 & 377 & 19.2\% & 17.1\%\\
    \hspace{3mm}\$25,000 to \$34,999 & 261 & 13.3\% & 8.3\%\\
    \hspace{3mm}\$35,000 to \$49,999 & 290 & 14.8\% & 11.7\%\\
    \hspace{3mm}\$50,000 to \$74,999 & 365 & 18.6\% & 16.5\%\\
    \hspace{3mm}\$75,000 to \$99,999 & 264 & 13.4\% & 12.3\%\\
    \hspace{3mm}\$100,000 to \$149,999 & 236 & 12.0\% & 15.5\%\\
    \hspace{3mm}\$150,000 or more & 170 & 8.7\% & 18.6\%\\
    \bottomrule
    \end{tabular}
    \begin{tablenotes}[para,flushright]
    \item[1] For gender, our source data from U.S.\ Census only have female and male percentage.\\
    \item[2] For Race, our source data from U.S.\ Census doesn't have Middle Eastern as a separate race.
    \end{tablenotes}
    \end{threeparttable}}
    \label{tab:sample_overview}
\end{table}

\subsubsection{Estimates of Adoption Rate}

For questions measuring the the five aspects of adoption (Section~\ref{sec:dependent_vars}), we grouped the options ``Somewhat agree'', ``Agree'' or ``Strongly agree'' to estimate the percentage of people that would install and use contact-tracing apps.
Table~\ref{tab:adoption_rate_estimates} summarizes the results.
58.9\% participants reported they at least somewhat agreed that they would install the app, which is close to findings of previous studies with U.S. smartphone users such as 55\% in~\citet{li2020decentralized} and 59\% in a national poll~\cite{Washingt23:online}

When participants were asked about actions they would take if they had installed the app, 76.2\% reported they at least somewhat agreed to report to the app if they tested positive for COVID-19.
Note that this is higher than the estimated install rate, which suggests that there are people who do not want to be tracked in general, but are less concerned to share the same information if they are infected to facilitate contact tracing.

Then we estimated the install retention rate in the long run in three different situations.
The \textit{Fewer cases} situation achieved the highest retention rate (63.7\%) and the \textit{Vaccine} situation achieved the lowest retention rate (57.6\%) which is similar to our expectation.
Although it is surprising to see more than half of the participants rated they would keep the app installed even when a vaccine becomes widely available.
This may be because some people have disbelief in vaccines or because they do not find these apps a big threat and tend not to actively uninstall the app once it is installed.
We also note that the install retention rate if the app drains the battery quickly (58.8\%) is close to the \textit{Vaccine} situation.
This suggests that practical concerns such as the impact on battery life can have crucial influence on users' decisions, which echos findings of prior work~\cite{redmilesuser}.

\begin{table}[htbp]
    \caption{Estimates of adoption rate (\%). A participant is considered as likely to install and use the app if choosing ``Somewhat agree'', ``Agree'', or ``Strongly agree'' for the corresponding statement (presented in Section~\ref{sec:dependent_vars}). The first column of each variable is the reference condition, and the conditions that have significantly different adoption intentions in our linear regression analyses in Table~\ref{tab:linear_regression_results} are marked in bold.}
    \vspace{0.5em}
    \resizebox{\linewidth}{!}{%
    \begin{threeparttable}
    \centering
    \begin{tabular}{p{0.15\linewidth} | p{0.05\linewidth}| p{0.09\linewidth}p{0.07\linewidth}p{0.06\linewidth} | p{0.05\linewidth}p{0.07\linewidth}p{0.078\linewidth} | p{0.05\linewidth}p{0.04\linewidth}p{0.04\linewidth}p{0.07\linewidth} | p{0.05\linewidth}p{0.05\linewidth}p{0.05\linewidth}p{0.08\linewidth}}
    \toprule
    Dependent & \multirow{2}{*}{All} & \multicolumn{3}{c|}{Proximity} & \multicolumn{3}{c|}{Location Use} & \multicolumn{4}{c|}{App Provider} & \multicolumn{4}{c}{Security risk}  \\ \cline{3-16}
                  variable &    & Decen.               &    Ano.C.   &  Id.C.     & None & Local & Upl. & State & Fed. & Tech & Empl. & None & Brea. & 2nd. & Re-id. \\ \midrule
    \multicolumn{16}{c}{\% of participants who agreed \textbf{``I will install this app if it becomes available in my area.''}}\\ \midrule
    Install & 58.9 & 58.4 & 60.1 & 58.4 & 57.1 & 58.7 & \textbf{61.1}** & 58.9 & 62.0 & 56.5 & 58.3 & 60.1 & 59.4 & \textbf{55.3}* & 60.0  \\ \midrule 
    \multicolumn{16}{c}{\% of participants who agreed \textbf{``I will report to this app if I test positive.''}}\\ \midrule
    Report & 76.2 & 74.6 & 76.1 & 78.0 & 76.0 & 76.0 & 76.6 & 75.5 & 78.4 & 75.0 & 75.9 & 76.9 & 76.5 & 76.7 & 74.5  \\ \midrule
    \multicolumn{16}{C{18cm}}{\% of participants who agreed \textbf{``I will keep this app installed} even if [my phone \textbf{battery} seems to last less long/COVID-19 \textbf{cases are steadily decreasing} in my area/a COVID-19 \textbf{vaccine} becomes widely available]''}\\ \midrule
    Battery & 58.8 & 58.4 & 60.3 & 57.8 & 58.5 & 58.5 & \textbf{59.4}* & 58.6 & 59.5 & 57.9 & 59.1 & 62.5 & 58.0 & 55.8	& 58.7  \\ \hline
    Fewer cases & 63.7 & 62.8 & 66.2 & 62.2 & 63.8 & 63.9 & 63.4 & 64.0 & 65.6 & 63.2 & 61.9 & 66.1 & 64.9 & 59.5 & 64.2  \\ \hline
    Vaccine & 57.6  & 57.3 & 58.5 & 57.2 & 56.0 & \textbf{58.4}* & \textbf{58.6}* & 58.6 & 58.9 & 58.4 & 54.6 & 60.4 & 57.8 & 55.1 & 57.1  \\ 
    \bottomrule
    \end{tabular}
    \begin{tablenotes}[para,flushright]
    \item[1]* p$<$ 0.05; ** p$<$0.01;\\
    \item[2] Condition names are abbreviated. Decen.: Decentralized; Ano.C.: Anonymized Centralized; Id.C.: Identified Centralized; None: No location use; Local: Location on device; Upl.: Location uploaded; State: State health authorities; Fed.: Federal health authorities; Tech: Tech company; Empl.: Employer or school; None: No security risk; Brea.: Data breach risk; 2nd.: Secondary use risk; Re-id.: Re-identification risk
    \end{tablenotes}
    \label{tab:adoption_rate_estimates}
    \end{threeparttable}}
\end{table}

\begin{table}[htbp]
    \caption{Estimates of percentage of people who at least somewhat agreed the app has security and privacy risks/self benefits/societal benefits and other people will install this app (\%). The first column of each variable is the reference condition.}
    \vspace{0.5em}
    \resizebox{\linewidth}{!}{%
    \begin{threeparttable}
    \centering
    \begin{tabular}{p{0.25\linewidth} | p{0.05\linewidth}| p{0.09\linewidth}p{0.07\linewidth}p{0.06\linewidth} | p{0.05\linewidth}p{0.07\linewidth}p{0.078\linewidth} | p{0.05\linewidth}p{0.04\linewidth}p{0.04\linewidth}p{0.07\linewidth} | p{0.05\linewidth}p{0.05\linewidth}p{0.05\linewidth}p{0.08\linewidth}}
    \toprule
    Mediator & \multirow{2}{*}{All} & \multicolumn{3}{c|}{Proximity} & \multicolumn{3}{c|}{Location Use} & \multicolumn{4}{c|}{App Provider} & \multicolumn{4}{c}{Security risk}  \\ \cline{3-16}
                  variable &    & Decen.               &    Ano.C.   &  Id.C.     & None & Local & Upl. & State & Fed. & Tech & Empl. & None & Brea. & 2nd. & Re-id. \\ \midrule
    \multicolumn{16}{c}{\% of participants who agreed \textbf{``Installing this app presents a risk to my security and privacy.''}}\\ \midrule
    S\&P risks & 54.8 & 52.7 & 52.5 & 59.1 & 50.2 & 57.1 & 57.3 & 54.1 & 50.4 & 56.8 & 58.1 & 43.3 & 62.8 & 59.7 & 53.6  \\ \midrule 
    \multicolumn{16}{c}{\% of participants who agreed \textbf{``Installing this app helps me protect myself against COVID-19.''}}\\ \midrule
    Self benefits & 68.2 & 68.6 & 67.0 & 69.0 & 66.2 & 67.7 & 70.8 & 70.2 & 70.9 & 65.4 & 66.4 & 69.8 & 67.1 & 67.2 & 68.9  \\ \midrule
    \multicolumn{16}{c}{\% of participants who agreed \textbf{``This app helps slow the spread of COVID-19 in my area.''}}\\ \midrule
    Societal benefits & 64.9 & 63.6 & 65.2 & 65.9 & 62.5 & 64.1 & 68.0 & 64.9 & 65.4 & 63.0 & 66.4 & 68.0 & 63.4 & 61.8 & 66.6  \\ \hline
    \multicolumn{16}{c}{\% of participants who agreed \textbf{``Most people in my area would install this app if it became available.''}}\\ \midrule
    Perceived adoption & 41.7 & 41.4 & 41.3 & 42.4 & 41.3 & 40.7 & 43.2 & 40.0 & 44.7 & 42.4 & 39.7 & 42.9 & 42.7 & 37.4 & 44.1  \\
    \bottomrule
    \end{tabular}
    \begin{tablenotes}[para,flushright]
    \item[1] Condition names are abbreviated. Decen.: Decentralized; Ano.C.: Anonymized Centralized; Id.C.: Identified Centralized; None: No location use; Local: Location on device; Upl.: Location uploaded; State: State health authorities; Fed.: Federal health authorities; Tech: Tech company; Empl.: Employer or school; None: No security risk; Brea.: Data breach risk; 2nd.: Secondary use risk; Re-id.: Re-identification risk
    \end{tablenotes}
    \label{tab:mediator_estimates}
    \end{threeparttable}}
\end{table}

\subsubsection{Estimates of Perceived Risks, Benefits, and Community Adoption Rate}

We also calculated estimates of the four mediator variables using the same method as in Table~\ref{tab:adoption_rate_estimates}. Table~\ref{tab:mediator_estimates} presents the result.
More people believed that installing the app could provide benefits to themselves (68.2\%) and to the society (64.9\%) than believed that installing the app would present a risk to their privacy and security (54.8\%).
Interestingly, only 41.7\% of our participants at least somewhat agreed that most people would install this app if it became available, which is much lower than the estimates of their own installation rate (58.9\%).
This suggests people generally hold an overly pessimistic attitude towards the adoption of contact-tracing app in the U.S.

These estimates also help validate the manipulations of our survey design.
For example, more people assigned to the \textit{identified centralized architecture} condition perceived security and privacy risks than people assigned to the \textit{decentralized architecture} condition (59.1\% vs. 52.7\%);
more people assigned to the two conditions that collect location data perceived security and privacy risks than people assigned to the \textit{no location use} condition (57.1\% and 57.3\% vs. 50.2\%).
More people assigned to the conditions that present one of the three security risks perceived security and privacy risks in the app than people assigned to the \textit{no security risk} condition (data breach risk: 62.8\%, secondary data use risk: 59.7\%, re-identification risk: 53.6\% vs. not priming about risk: 43.3\%).
These results are in line with our expectations when designing the conditions, which demonstrates that our between-subjects design effectively conveyed the key characteristics of the app and our participants were able to correctly understand these characteristics before reporting their subjective feelings.

\subsection{Effects of App Design Choices on Adoption Intentions (RQ1)}
\label{sec:RQ1_results}

In RQ1, we are interested in investigating how app design choices such as decentralized vs. centralized architecture, location use, app providers, and the description about security risks in the app affect one's adoption intentions.
According to the linear regression results demonstrated in Table~\ref{tab:linear_regression_results}, location use of the app (RQ1.2) and the disclosure of the secondary data use security risk (RQ1.4) had significant effects on several aspects of adoption intentions.
Conversely, the difference in decentralized vs. centralized architectures (RQ1.1) and app providers (RQ1.3) did not have a significant effect on adoption intentions.
We calculated the $f^2$ scores of factors related to app design choices to measure their effect size for the five outcome variables.
The $f^2$ for the five outcome variables are 0.006, 0.004, 0.003, 0.002, 0.004 respectively, which shows app design choices have a very small effect on adoption intentions in general\footnote{\label{note:f-squared-rule-of-thumb}The basic rule of thumb to interpret $f^2$ is: $f^2=0.02$ indicates a small effect; $f^2=0.15$ indicates a medium effect; $f^2=0.35$ indicates a large effect~\cite{cohen2013statistical}.}.

For location use, we can see that the condition \textit{Location on device} and \textit{Location uploaded} both had positive significant effects on some aspects of adoption intentions.
For example, the coefficient for \textit{Location uploaded} condition for the \textit{Install app} outcome variable is 0.258, which represents an estimated increase in the 7-point installation intention rating if the app collects location data and uploads data to the servers for analyzing infection hotspots as compared to not providing location features at all.
This suggests contributing a little more location data in exchange for more useful information could drive more adoption of the app.
We want to note that the two location features follow privacy-by-design principles by only storing and uploading the data if necessary (e.g., only uploading location data of users who test positive rather than all users).
These design considerations should also be taken into consideration when interpreting the positive effects of these features.

For security risk, although prompting participants about all three types of risks consistently increased their perceptions about the risks to their security and privacy caused by installing this app (Table~\ref{tab:mediator_estimates}), only showing the secondary data use risk significantly decreased the installation intentions.
This suggests that one's security and privacy concerns may not be a determinant of their adoption intentions, which is further supported with our mediation analysis (Section~\ref{sec:mediation_results}).

In addition, the difference between the reactions to the secondary data use risk and the other two types of risks provides us with another aspect to view the comparisons between decentralized and centralized architectures.
As centralized architecture requires more data to be stored on central servers, app users will become more vulnerable to the data breach risk and secondary data use risk than decentralized architectures.
Therefore, although there is no significant difference in people's adoption intentions when directly comparing the two architectures, our results suggest that using a decentralized architecture could help reduce security risks that people are more concerned about.


\begin{table}[htbp]
    \caption{Linear regression results: The main effects of app design choices and individual differences on app adoption intentions. As described in Section~\ref{sec:independent_vars}, we excluded the data from groups that contained too few responses (e.g., Non-binary gender, Native Hawaiian or Pacific Islander). The sample used for the regression analysis contains \samplesizecleaned{} responses.}
    \vspace{0.5em}
    \resizebox{\linewidth}{!}{%
    \begin{threeparttable}
    \centering
    \begin{tabular}{p{0.4\linewidth}   C{0.25\linewidth}C{0.25\linewidth}C{0.25\linewidth}C{0.25\linewidth}C{0.25\linewidth}}
    \toprule
    \multirow{3}{*}{Independent variable} & \multirow{2}{*}{Install app} & \multirow{2}{*}{Report positive} &
    \multicolumn{3}{c}{Install retention} \\
     & & & Battery & Fewer cases & Vaccine\\
        & Coef. (S.E.) & Coef. (S.E.)& Coef. (S.E.) & Coef. (S.E.)& Coef. (S.E.)\\\midrule
    (Intercept) & -1.663*** (0.428) & -0.531 (0.413) & -2.129*** (0.446) & -1.968*** (0.432) & -2.296*** (0.451) \\
    \textbf{Factors related to app design choices} \\
    Proximity (Decentralized=0) \\
    \hspace{3mm} Anonymized Centralized & -0.067 (0.092) & 0.077 (0.089) & 0.002 (0.096) & -0.014 (0.093) & -0.062 (0.097) \\
    \hspace{3mm} Identified Centralized & -0.036 (0.091) & 0.110 (0.087) & -0.002 (0.094) & -0.017 (0.091) & -0.038 (0.095)\\
    Location use (No use=0)\\
    \hspace{3mm} Location on device & 0.098 (0.092) & 0.065 (0.089) & 0.086 (0.096) & 0.127 (0.0926) & 0.210* (0.097)\\
    \hspace{3mm} Location uploaded & 0.258** (0.091) & 0.119 (0.088) & 0.189* (0.095)& 0.072 (0.092) & 0.203* (0.096)\\
    App provider (State=0) \\
    \hspace{3mm} Federal health authorities  & 0.132 (0.106) & 0.167 (0.102) & 0.029 (0.110) & 0.084 (0.107)& 0.050 (0.111)\\
    \hspace{3mm} Tech company & -0.050 (0.105) & -0.032 (0.101) & -0.038 (0.109) & -0.045 (0.106)&-0.046 (0.111)\\
    \hspace{3mm} Employer or school & -0.008 (0.106) & 0.121 (0.103) & 0.091 (0.111)& -0.005 (0.107) & -0.074 (0.112)\\
    Security risk (No risk=0)  \\
    \hspace{3mm} Data breach risk & -0.067 (0.104) & -0.015 (0.100) & -0.039 (0.108) & -0.053 (0.105) & -0.099 (0.109)\\
    \hspace{3mm} Secondary use risk & -0.209* (0.104) & -0.046 (0.101)  & -0.039  (0.108) & -0.140 (0.106) & -0.197 (0.110)\\
    \hspace{3mm} Re-identification risk & -0.141 (0.107) & -0.099 (0.104) & -0.090 (0.112) & -0.085 (0.108) & -0.148 (0.113)\\
    \textbf{Factors related to individual differences} \\ 
    Prosocialness & 0.418*** (0.053) & 0.298*** (0.051) & 0.475*** (0.055) & 0.444*** (0.053) &  0.544*** (0.056)\\
    General privacy concern & -0.113* (0.047) & -0.003 (0.045) & -0.131** (0.049) & -0.080 (0.047) & -0.125* (0.049)\\
    COVID-19 risk perception & 0.682*** (0.047) &  0.681*** (0.046) & 0.659*** (0.049) & 0.718*** (0.048) & 0.697*** (0.050)\\
    Technology readiness & 0.689*** (0.067) & 0.561*** (0.064) & 0.607*** (0.070) & 0.623*** (0.067) & 0.596*** (0.070)\\
    Age & -0.011*** (0.003) & 0.002 (0.002) & 0.008** (0.003) & 0.001 (0.003) &  0.002 (0.003)\\
    Gender (Male=0) \\
    \hspace{3mm} Female & -0.362*** (0.082) & -0.184* (0.079) & -0.402*** (0.085) & -0.264** (0.082) & -0.243** (0.086)\\
    Race (White=0) \\
    \hspace{3mm} Asian & 0.261 (0.155) & 0.256 (0.150) & 0.287 (0.161) & 0.241 (0.156) &  0.415* (0.163)\\
    \hspace{3mm} Black/African American & 0.148 (0.119) & 0.114 (0.115) & 0.220 (0.124) & 0.356** (0.120) &  0.512*** (0.125)\\
    \hspace{3mm} Hispanic/Latino & 0.276* (0.120) & 0.231* (0.116) & 0.066 (0.125) & 0.346** (0.121) &  0.399** (0.127)\\
    Education & 0.034 (0.023) & 0.011 (0.022) & 0.036 (0.024) & 0.048* (0.023) &  0.031 (0.024)\\
    Household Income & 0.110*** (0.023) & 0.040 (0.023) & 0.103*** (0.024) & 0.059* (0.024) &  0.057* (0.025)\\
    Essential worker & -0.203* (0.098) & -0.316*** (0.094) & -0.225* (0.102) & -0.167  (0.098) & -0.120 (0.103)\\
    Health worker & -0.013 (0.151) & -0.202 (0.146) & 0.045 (0.158) & -0.073  (0.153) & -0.007 (0.159)\\
    Public transit use & 0.155*** (0.036) & 0.051 (0.035) & 0.205*** (0.038) & 0.129*** (0.037) &  0.219*** (0.038)\\
    Urban area percentage & 0.006** (0.002) & 0.002 (0.002) & 0.003 (0.002) & 0.003 (0.002) &  0.004* (0.002)\\\midrule
    $R^2$ & 0.329 & 0.226 & 0.278 & 0.282&0.289\\
    Adjusted $R^2$  & 0.320 & 0.216 & 0.269 & 0.272& 0.279\\
    \bottomrule
    \end{tabular}
    \begin{tablenotes}[para,flushright]
    Note: * p$<$0.05; ** p$<$0.01; *** p$<$0.001
    \end{tablenotes}
    \label{tab:linear_regression_results}
    \end{threeparttable}}
\end{table}

\subsection{Effects of Individual Differences on Adoption Intentions (RQ2)}
\label{sec:RQ2_results}

\subsubsection{Main Effects (RQ2.1-2.6)}
\label{sec:individual_difference_main_effect_results}
In RQ2, we are interested in investigating how individual differences such as demographic factors and other personal characteristics like prosocialness and general privacy concerns affect one's adoption intentions.
Table~\ref{tab:linear_regression_results} presents the results.
The $f^2$ of all factors related to individual differences for the five outcome variables are 0.475, 0.286, 0.380, 0.385, 0.397 respectively, which shows that factors related to individual differences have a very large effect on adoption intentions, especially app installation intentions.






We found prosocialness, COVID-19 risk perception, and technology readiness all had significant positive effect on the five aspects of adoption intentions.
Conversely, general privacy concern had a significant negative effect on three out of the five aspects of adoption intentions.
These results are consistent with our expectations and answer RQ2.1-2.4.

We found multiple demographic factors had significant effects on adoptions intentions (RQ2.5).
Females had significantly lower intentions to adopt the app in all five aspects, especially for installing the app (Coef. = -0.362, i.e., our model predicts females' 7-point installation intention rating 0.362 lower than males).
High household income had significant positive effects on intentions to install the app and keep the app installed but had no significant effect on intentions to report positive cases.
Higher education had significant positive effects on intentions to keep the app installed when the COVID-19 cases are steadily decreasing.
Older people had significantly lower intentions to install the app, while they had significantly higher intentions to keep the app installed when the battery seems to last less long.

Unlike other demographic factors, the significant effects of race mostly appeared on the intentions to keep the app installed rather than the intentions to install the app.
For example, although only Hispanics had significantly higher intentions to install the app than Whites, Asians, Blacks, and Hispanics all had significantly higher intentions to keep the app installed even if a vaccine becomes widely available.
Note that the causes of the higher intentions to keep the app installed for the three races could be different.
For example, Pew research recently found that Black Americans are less inclined to get vaccinated than other racial and ethnic groups and Asians are the most inclined to get vaccinated~\cite{Intentto82:online}.
This requires further investigation by future work.

For people who are at higher risks to get exposed to COVID-19, we found people who are frequent public transit users during the pandemic and people who live in more urbanized areas had significantly higher adoption intentions.
To our surprise, essential workers had significantly lower adoption intentions in several aspects, especially for reporting positive cases to the app (Coef.=-0.316).
Our mediation analysis provides more insights into possible causes of this finding (Section~\ref{sec:mediation_results}).
Note that the average app installation intention rating for all essential workers (mean=4.77) is actually slightly higher than other participants (mean=4.54)
This may be because essential workers are generally younger (median age=35) than the rest of the sample (median age=49) and we have showed that younger people are more inclined to adopt contact-tracing apps, which counteracted the influence of being an essential worker.


\subsubsection{Interaction Effects (RQ2.7)}
\label{sec:individual_difference_interaction_effect_results}

\begin{figure}
    \centering
    \includegraphics[width=0.45\linewidth]{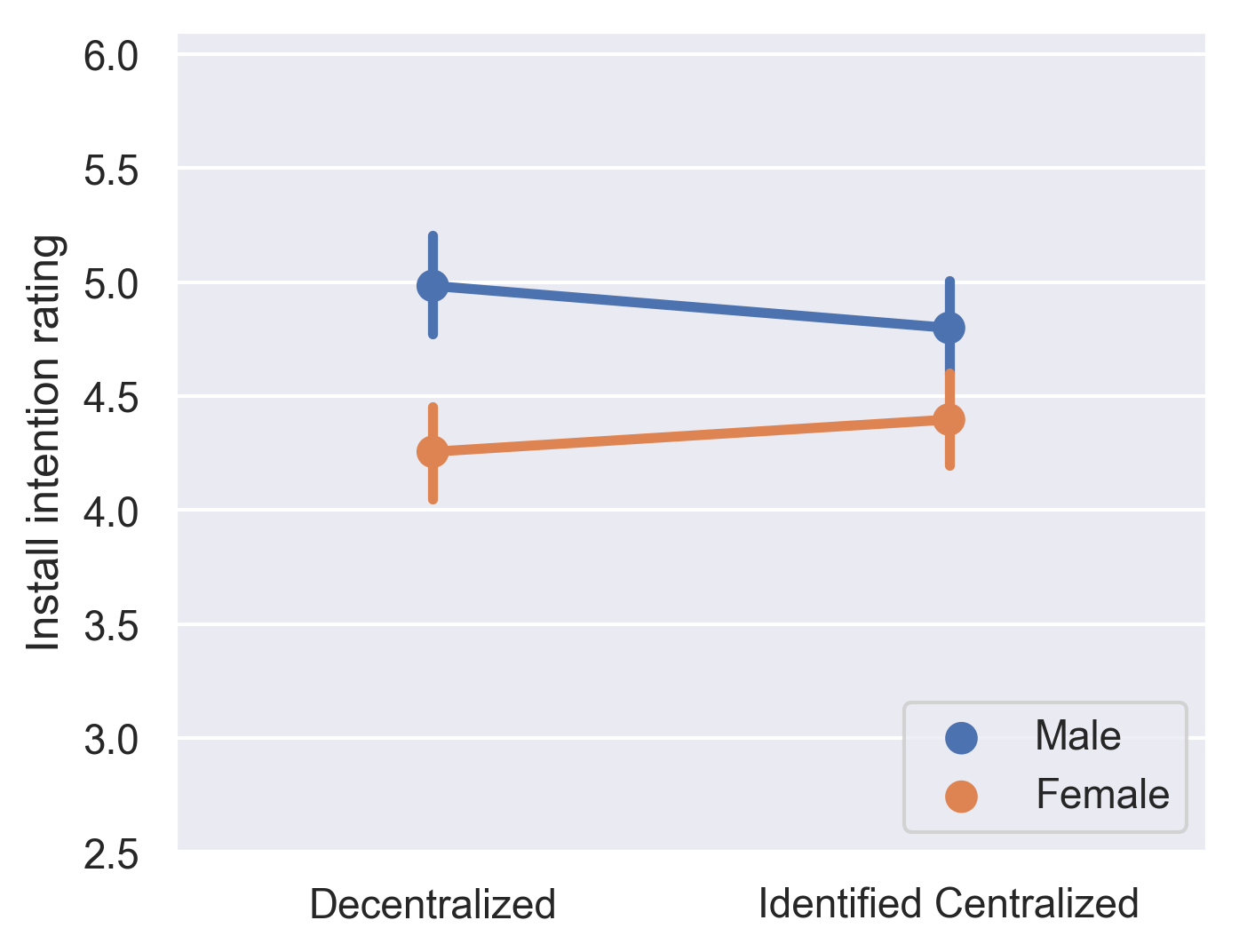}
    \includegraphics[width=0.45\linewidth]{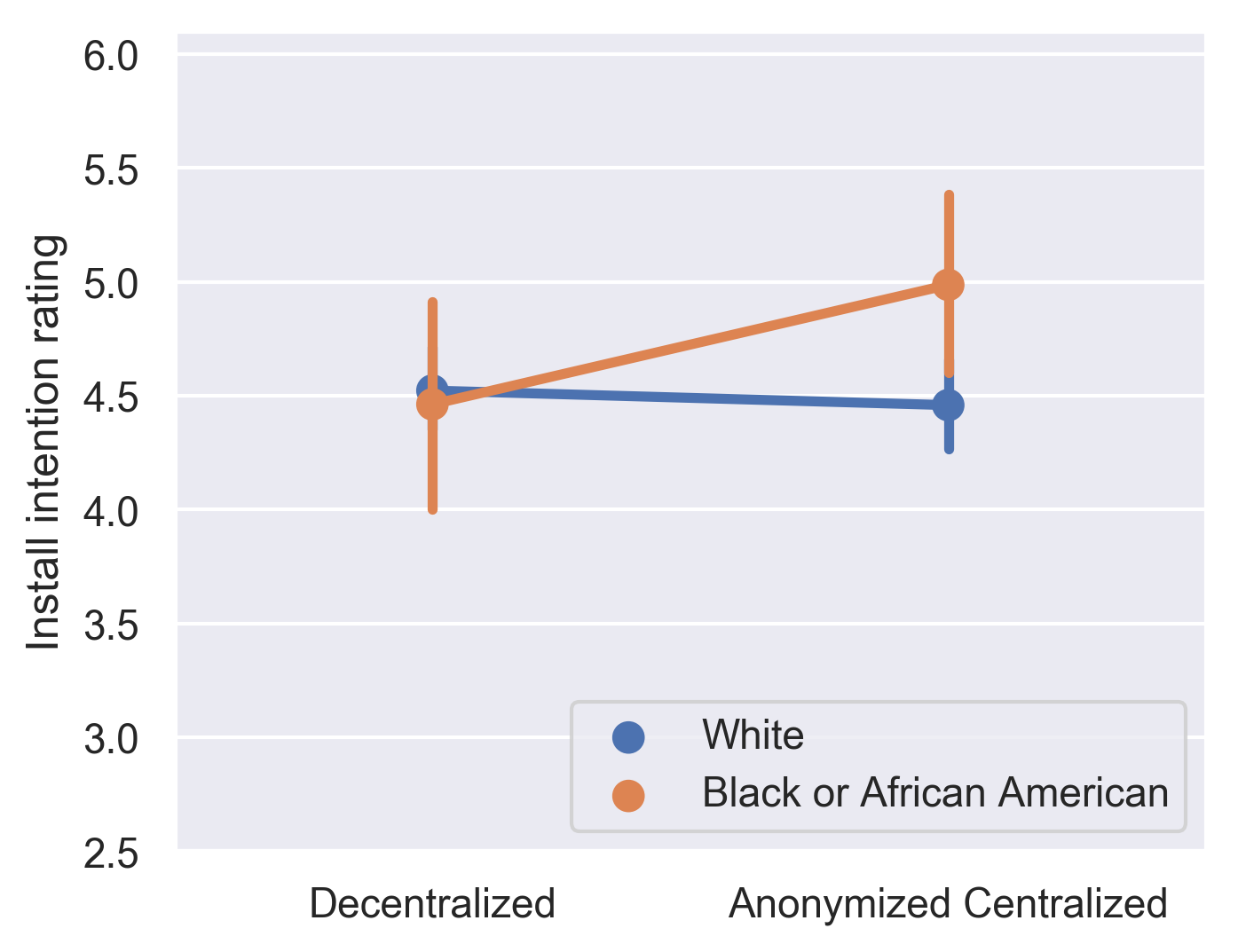}
    \includegraphics[width=0.45\linewidth]{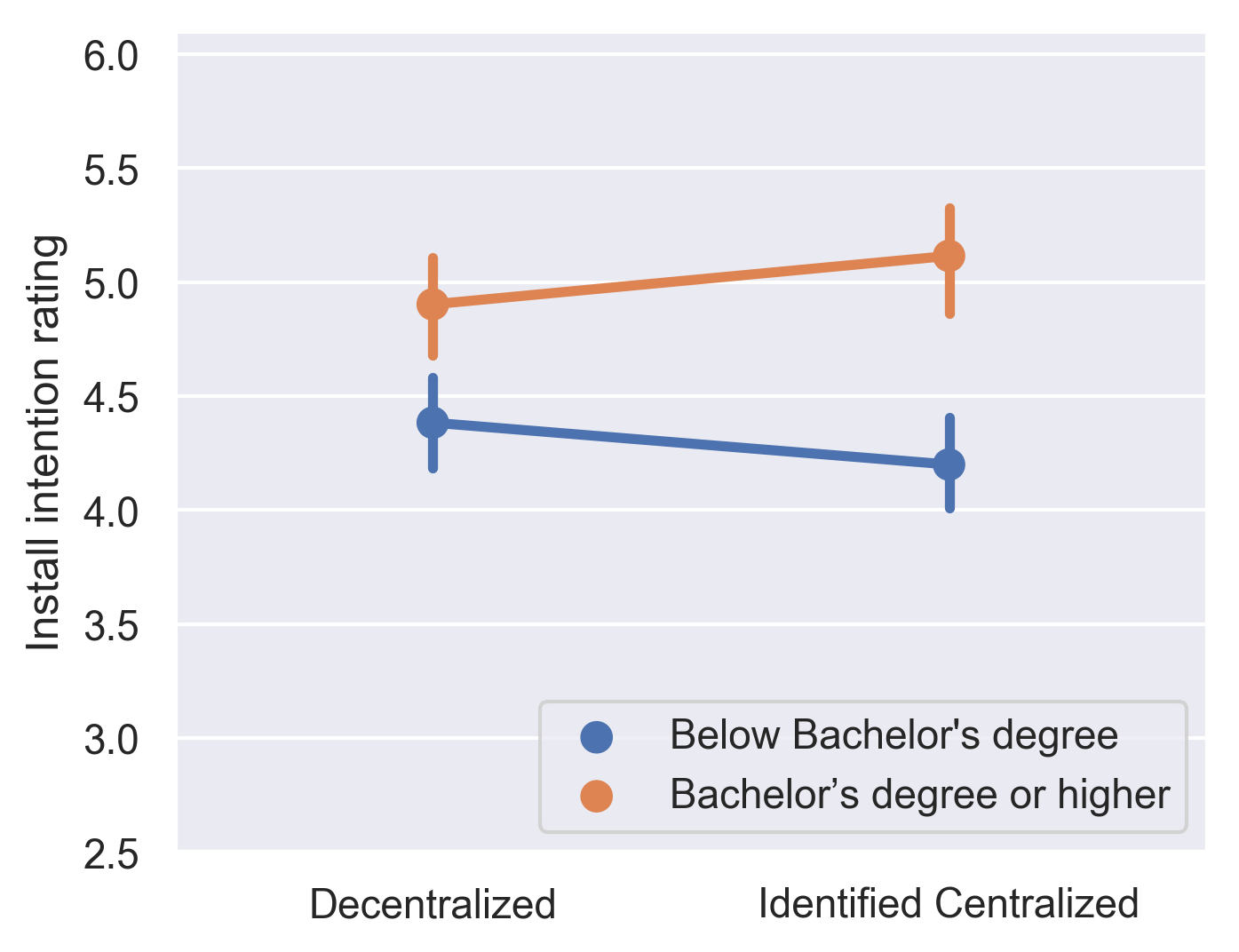}
    \caption{Significant interaction effects: Proximity-based contact tracing x Demographics. The vertical bars represent the estimated 95\% confidence intervals of the ``\textit{Install app}'' intention rating. Note that we group the eleven education levels into two classes for illustrative purposes.}
    \label{fig:interaction_effects_proximity_demographics}
\end{figure}

\begin{figure}
    \centering
    \includegraphics[width=0.45\linewidth]{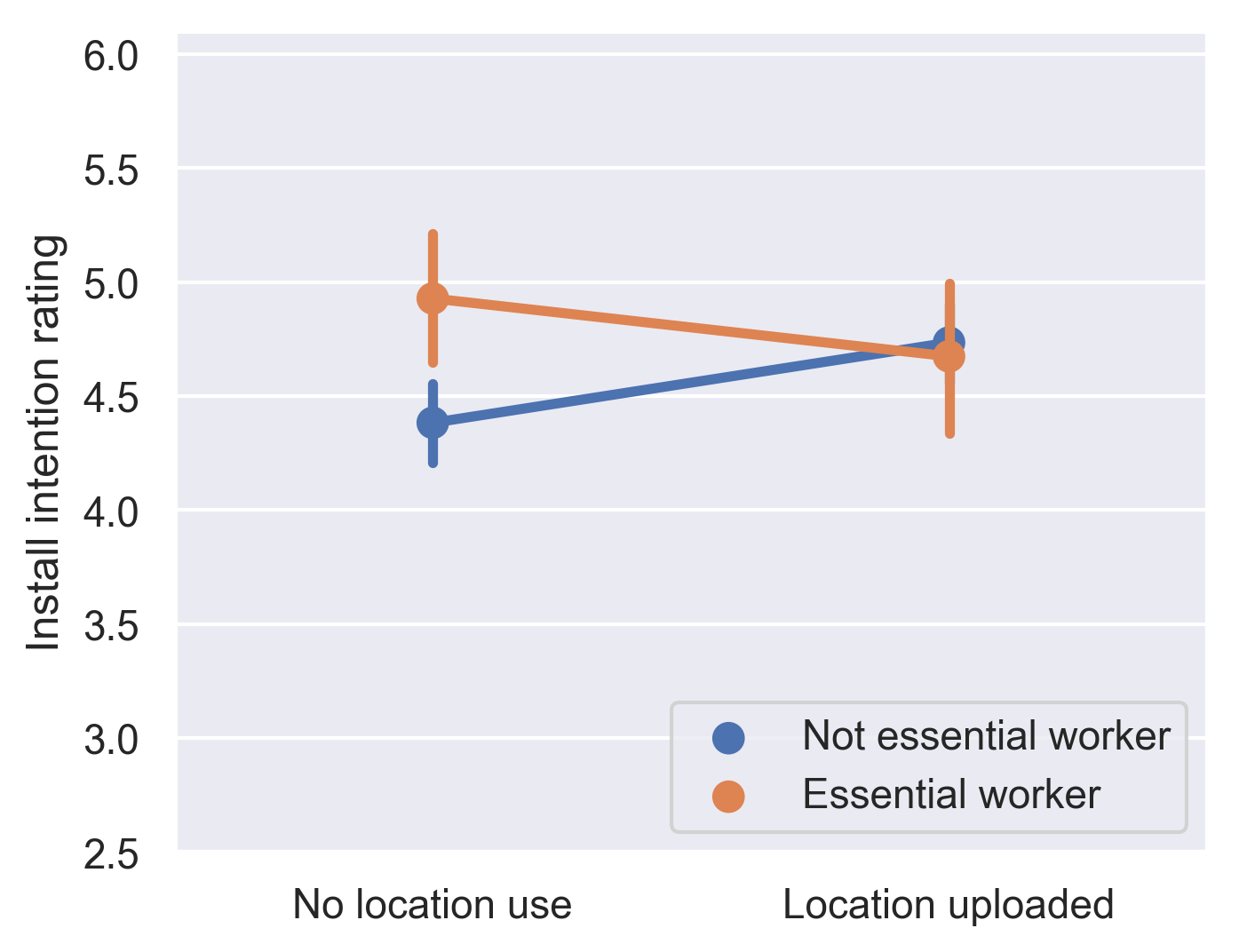}
    \includegraphics[width=0.45\linewidth]{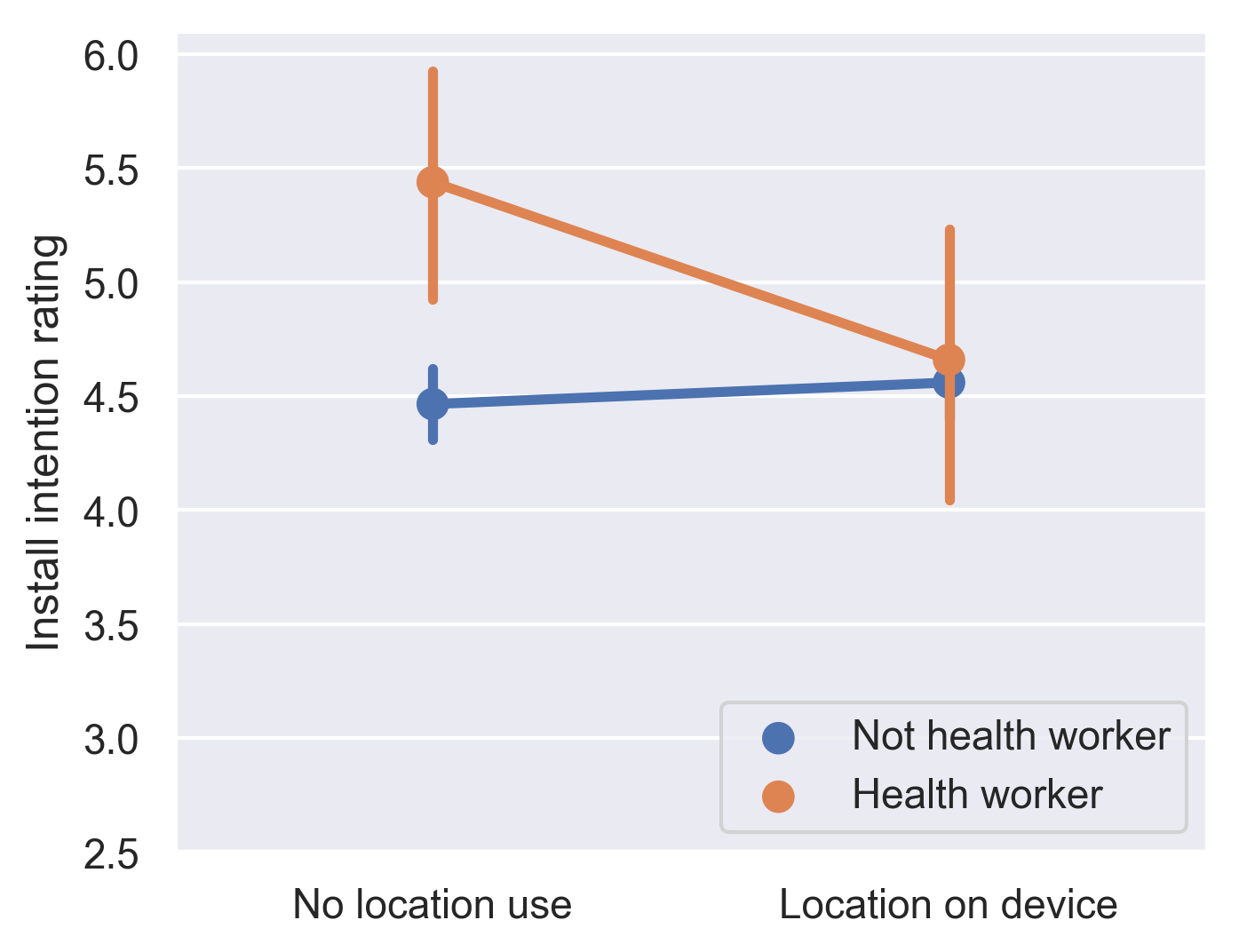}
    \caption{Significant interaction effects: Location use x Demographics. The vertical bars represent the estimated 95\% confidence intervals of the ``\textit{Install app}'' intention rating.}
    \label{fig:interaction_effects_location_demographics}
\end{figure}

\begin{figure}
    \centering
    \includegraphics[width=0.45\linewidth]{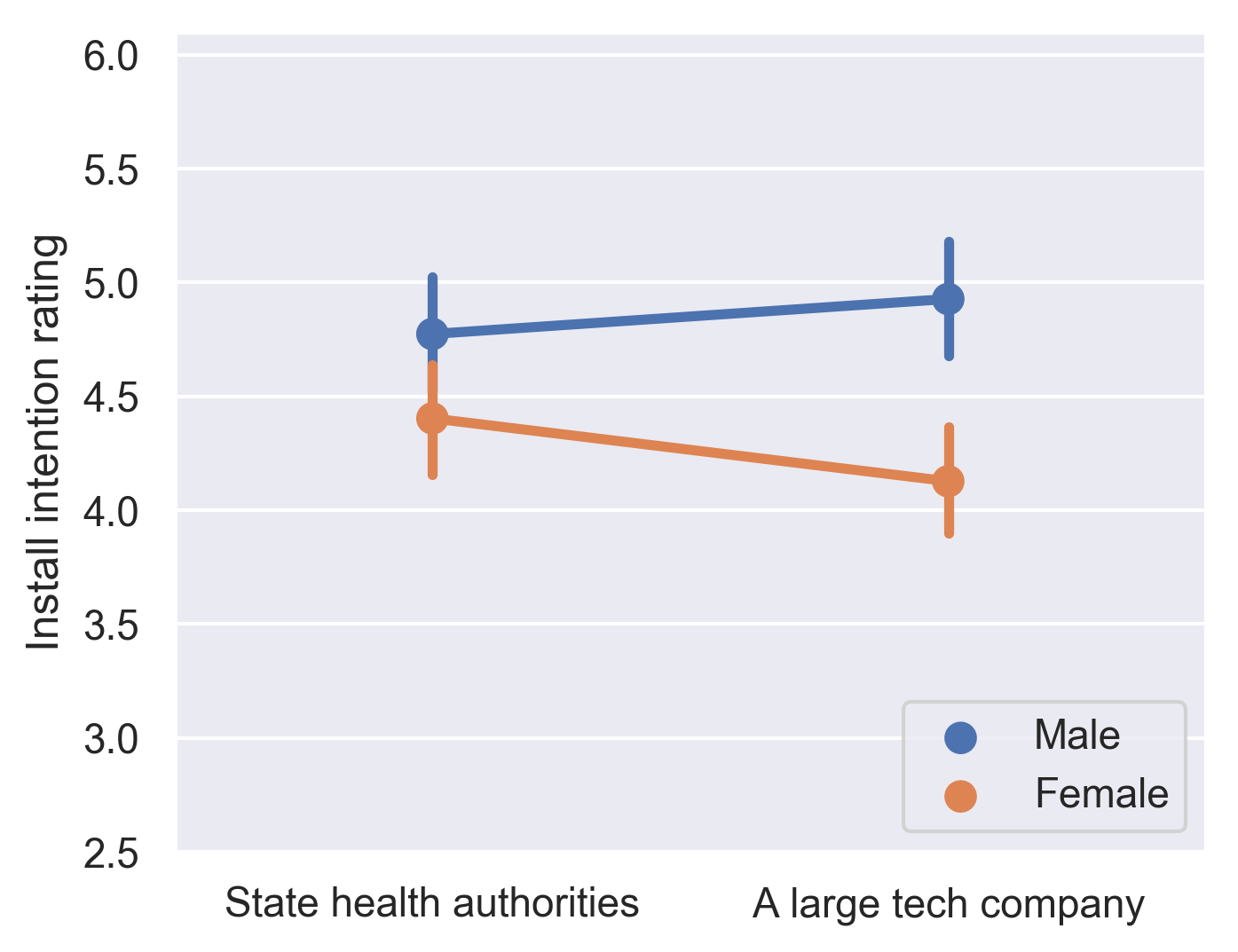}
    \includegraphics[width=0.45\linewidth]{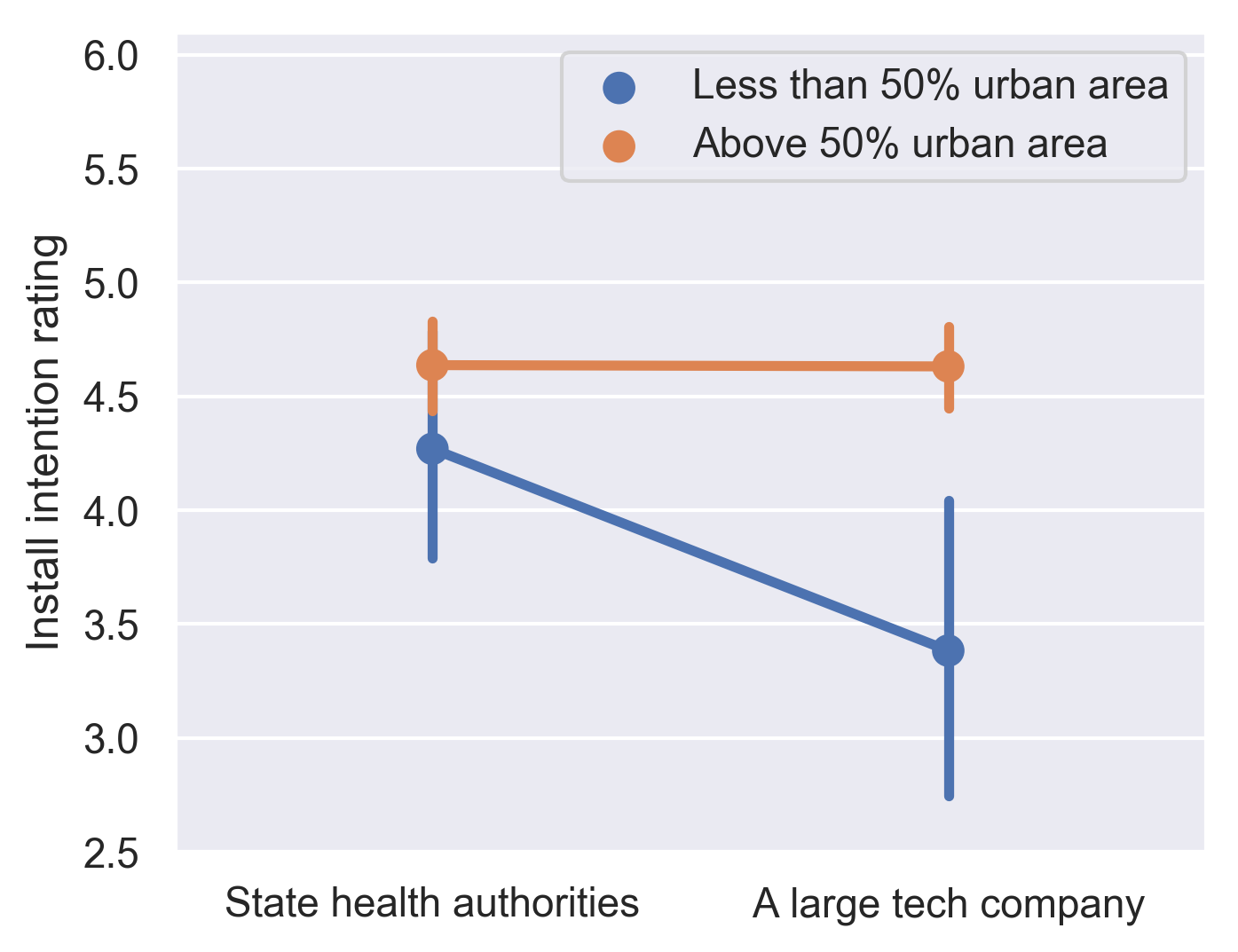}
    \caption{Significant interaction effects: App provider x Demographics. The vertical bars represent the estimated 95\% confidence intervals of the ``\textit{Install app}'' intention rating. Note that we group the urban area percentage values into two classes for illustrative purposes.}
    \label{fig:interaction_effects_app_provider_demographics}
\end{figure}

In RQ2.7, we focus on the \textit{app installation intentions} and study if the same app design could result in different installation intentions for different sub-populations.
This could help us predict the adoption of a certain app design by people who are at different levels of risks of getting exposed to or infected with COVID-19 and analyze the implications of potentially unbalanced app adoption.
To answer this research question, we built a interactive model for the ``\textit{Install app}'' outcome variable which includes the interaction between factors related to app design choices and demographic factors.
Due to the space constraints, we only present the interaction that had a significant effect in Figure~\ref{fig:interaction_effects_proximity_demographics}, \ref{fig:interaction_effects_location_demographics}, \ref{fig:interaction_effects_app_provider_demographics}, \ref{fig:interaction_effects_risk_demographics}.
The complete results can be found in the Appendices (Table~\ref{tab:interaction_regression_results}).

For the interaction between proximity-based contact tracing design and demographic factors (Figure~\ref{fig:interaction_effects_proximity_demographics}), we found that the effects of different architectures to achieve proximity-based contact tracing are moderated by gender, race, and education level factors.
Specifically, females tended to prefer the identified centralized architecture while males tend to prefer the decentralized architecture (Coef.=0.624, p$<$.01).
The difference in installation intentions between Black and White people was exacerbated when changing from decentralized to anonymized centralized architecture (Coef.=0.662, p$<$.05).
People who received higher education preferred identified centralized architectures to the decentralized architecture (Coef.=0.118, p$<$.05).

For the interaction between location use and demographic factors (Figure~\ref{fig:interaction_effects_location_demographics}), we found that the effects of location use are moderated by whether the person is a essential/health worker. Although the ``\textit{Location uploaded}'' feature could drive a significantly higher installation intention rating at the population level, essential workers preferred the ``\textit{No location use}'' condition (Coef.=-0.544, p$<$.05).
Similarly, health workers preferred the ``\textit{No location use}'' condition a lot more than the ``\textit{Location on device}'' condition (Coef.=-0.939, p$<$.05).

For the interaction between app provider and demographic factors (Figure~\ref{fig:interaction_effects_app_provider_demographics}), we found that the effects of app provider are moderated by gender and urban area percentage.
Females (Coef.=-0.621, p$<$.01) and people living in less urbanized areas (Coef.=0.0119, p$<$.05) tended to prefer contact-tracing apps provided by the state health authorities to a large tech company.

\begin{figure}
    \centering
    \includegraphics[width=0.45\linewidth]{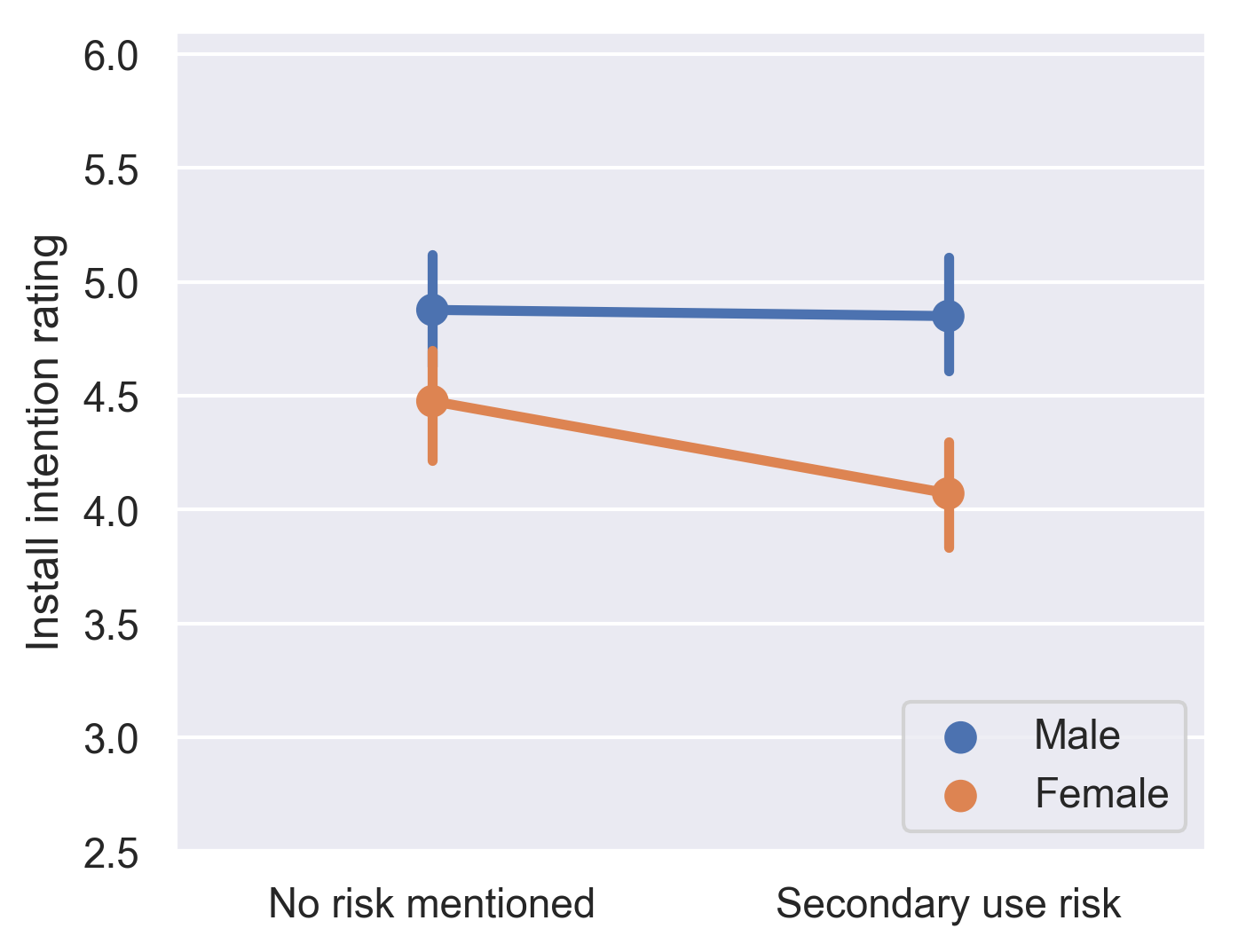}
    \includegraphics[width=0.45\linewidth]{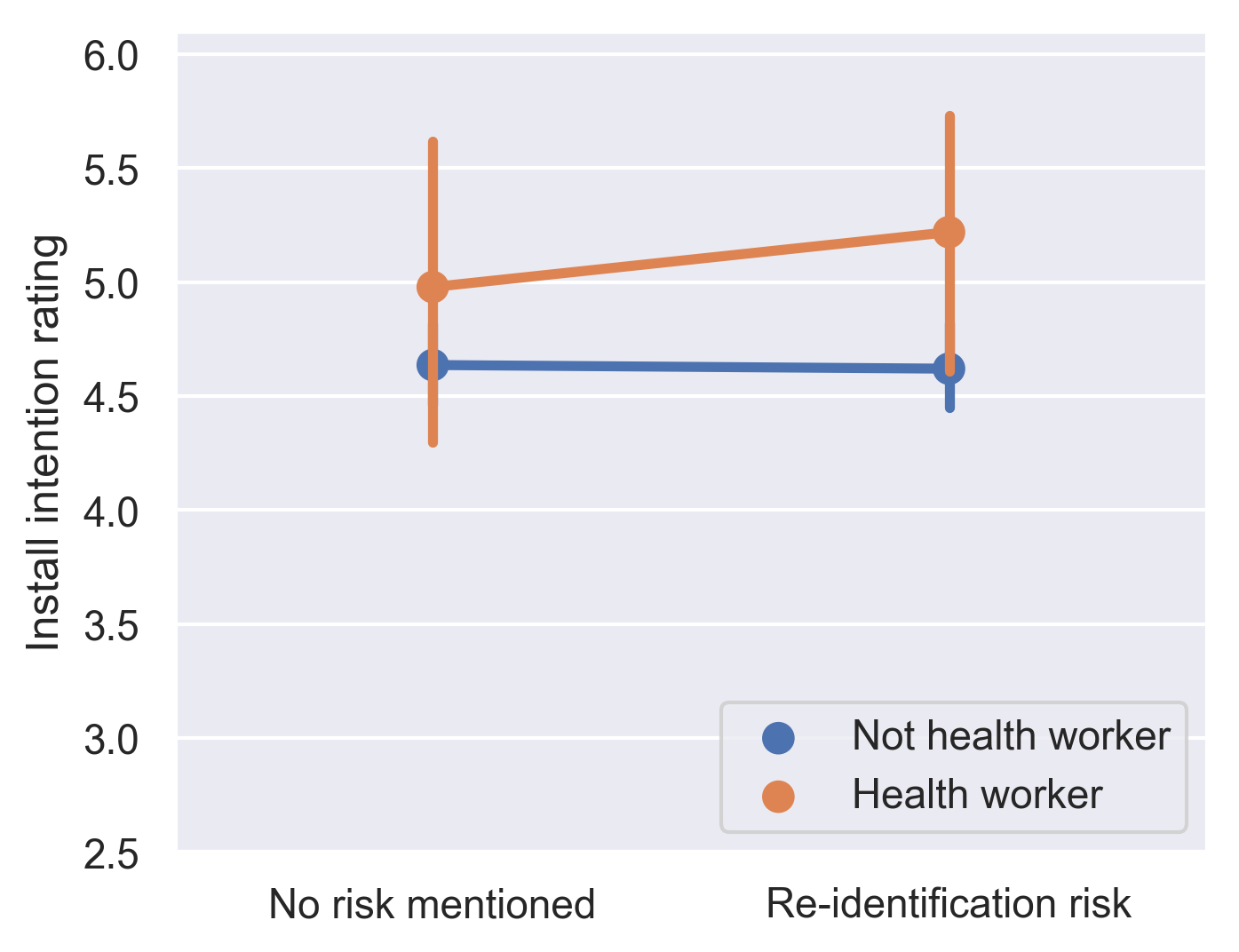}
    \includegraphics[width=0.45\linewidth]{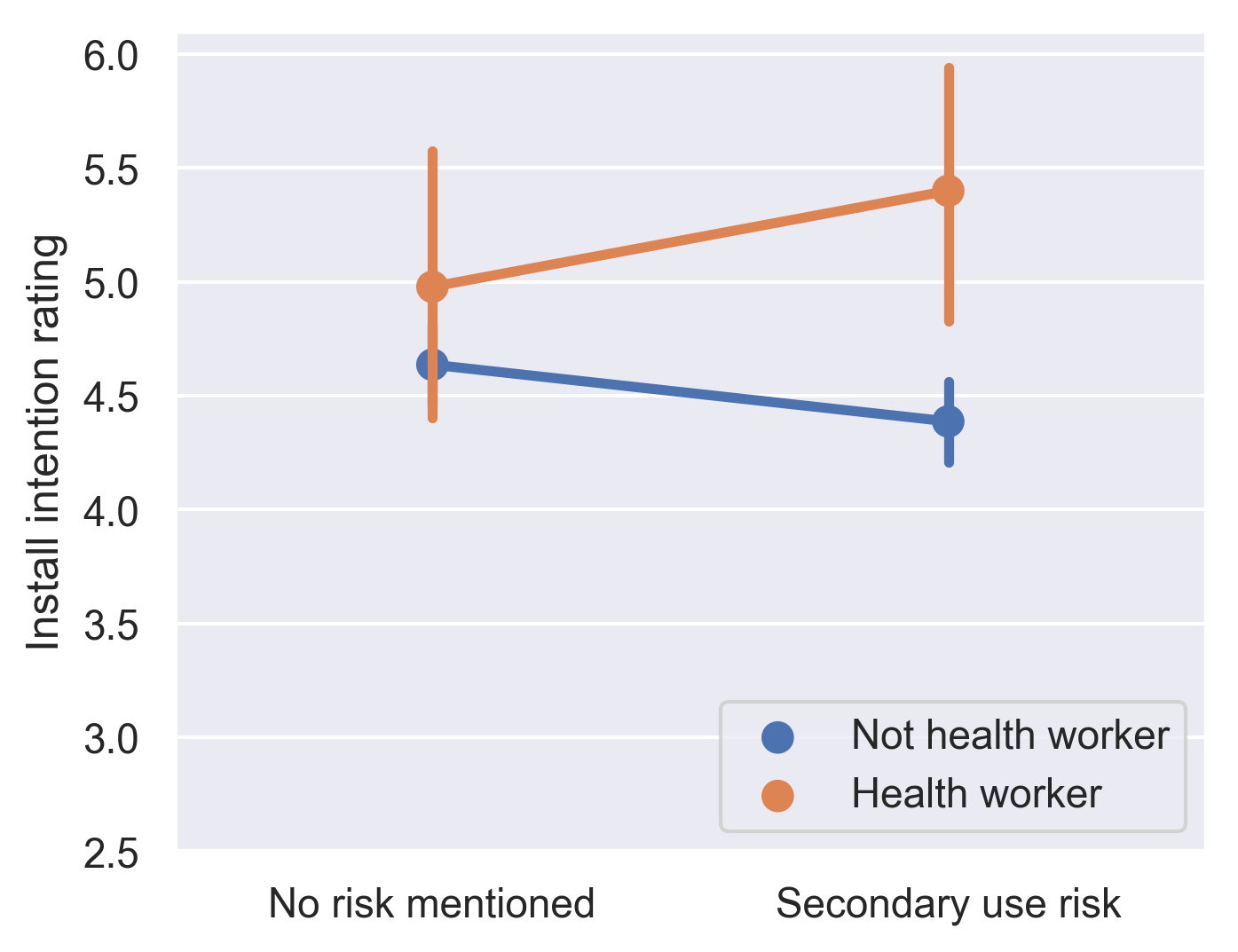}
    \caption{Significant interaction effects: Security risk presentation x Demographics. The vertical bars represent the estimated 95\% confidence intervals of the ``\textit{Install app}'' intention rating.}
    \label{fig:interaction_effects_risk_demographics}
\end{figure}

For the interaction between security risk presentation and demographic factors (Figure~\ref{fig:interaction_effects_risk_demographics}), we found that the effects of security risk are moderated by gender and whether the person is a health worker.
Specifically, females were more discouraged by the secondary data use risk than males (Coef.=-0.471, p$<$.05);
People who are not health workers were more discouraged by the secondary data use risk (Coef.=0.878, p$<$.05) and the re-identification risk than health workers (Coef.=1.05, p$<$.05).

\subsection{Explaining the Effects of App Design Choices and Individual Differences (RQ3)}
\label{sec:mediation_results}

\begin{figure}
    \centering
    \includegraphics[width=1\linewidth]{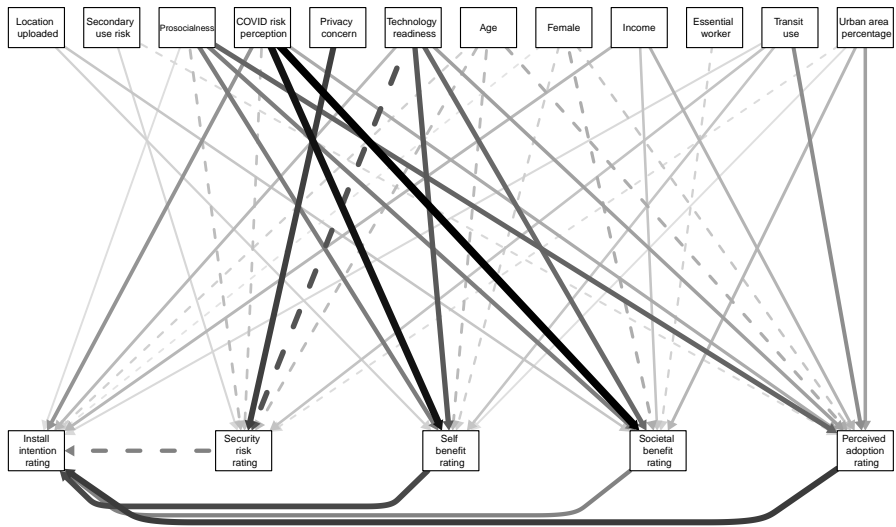}
    \caption{An illustration of mediation effects that explain how certain app design choices and individual differences affect people's intentions to install the app. Edges in \textbf{solid lines} (e.g., \textit{Secondary use risk} $\rightarrow$ \textit{Security risk rating}) indicate positive correlation and Edges in \textbf{dashed lines} (e.g., \textit{Technology readiness} $\rightarrow$ \textit{Security risk rating}) indicate negative correlation. Only edges that have \textit{significant effects} are plotted and the edge weight and transparency corresponds to the standardized coefficients (NOT effect size; effect size is presented in Table~\ref{tab:mediation_indirect_effects}). There is a significant indirect effect between an independent variable and the outcome variable (i.e., install intention rating) through a mediator variable if there is a pathway from the independent variable to the outcome variable through the mediator variable.}
    \label{fig:sem}
\end{figure}

In RQ3, we aim to explain how certain app design choices and individual differences had significant effects on one's app installation intentions through the four mediator variables: security and privacy risks, self benefits, societal benefits, and perceived adoption.
To answer this research question, we conducted a mediation analysis using structural equation modeling following methods of previous research~\cite{preacher2011effect}. We measured the relative magnitude of the indirect effects through the four mediators and calculated the 95\% confidence intervals of these ratios using a bootstrap approach.
Table~\ref{tab:mediation_indirect_effects} presents the ratios of the indirect effects to the total effects and their 95\% confidence intervals for each pair of independent variables and mediator variables.
Figure~\ref{fig:sem} illustrates the significant correlations between independent variables and mediator variables and between mediator variables and the outcome variable installation intention rating.
Our model fit is acceptable according to the Standardized Root Mean Square Residual (SRMR=0.057)\footnote{The rule-of-thumb to interpret SRMR is: SRMR less than 0.05 means the model fits well; SRMR less than 0.08 means the model fit is acceptable~\cite{hu1999cutoff,cangur2015comparison}.}

\begin{table}[]
    \centering
    \caption{This table shows the estimated ratios of the indirect effect to the total effect, which can be interpreted loosely as the percentages of the total effect of an independent variable on installation intention that are achieved through the four mediator variables. The cells that contain a ratio and its 95\% confidence interval indicate that the independent variable (e.g., ``\textit{Location uploaded}'') affects the app installation intentions through the mediator variable (e.g., ``\textit{Self benefits}'') and the indirect effect is significant. A positive number means the indirect effect is in the same direction as the total effect and a negative number means the indirect effect is in the opposite direction of the total effect which counteracts the other variables' positive effects. We leave the cell blank (``--'') if the indirect effect is not significant.}
    \vspace{0.5em}
    \resizebox{\linewidth}{!}{
    \begin{tabular}{p{0.35\linewidth} p{0.25\linewidth} p{0.25\linewidth} p{0.25\linewidth} p{0.25\linewidth}}
    \toprule
    Independent var.     & S\&P risks & Self benefits & Societal benefits & Perceived adoption \\
    \midrule
Location uploaded & -- & 0.21 [0.08, 0.56] & 0.18 [0.08, 0.48] & --\\
Secondary use risk & 0.17 [-0.05, 0.89] & -- & -- & 0.26 [-0.06, 1.26]\\
Prosocialness & 0.08 [0.04, 0.13] & 0.22 [0.15, 0.32] & 0.15 [0.09, 0.23] & 0.28 [0.20, 0.39]\\
COVID-19 risk perception & 0.04 [0.02,  0.07] & 0.24 [0.17,  0.31] & 0.18 [0.10, 0.25] & 0.09 [0.05,  0.13]\\
General privacy concern & 0.77 [0.36,  3.55] & -- & -- & --\\
Technology readiness & 0.15 [0.10,  0.22] & 0.22 [0.15,  0.30] & 0.13 [0.08,  0.20] & 0.13 [0.08,  0.19]\\
Age & -0.15 [-0.35, -0.05] & 0.21 [0.11,  0.37] & -- & 0.27 [0.15,  0.48]\\
Female & -- & 0.15 [0.06,  0.28] & 0.18 [0.09,  0.32] & 0.19 [0.09,  0.35]\\
Hispanic & -- & -- & -- & --\\
Income & -- & -- & 0.10 [0.04,  0.20] & 0.22 [0.12,  0.38]\\
Essential worker & -- & -- & 0.19 [0.04,  0.89] & --\\
Transit use & -0.13 [-0.33, -0.04] & 0.19 [0.09,  0.34] & -- & 0.39 [0.25,  0.70]\\
Urban area percentage & 0.11 [0.01,  0.29] & 0.13 [-0.00,  0.35] & 0.21 [0.10,  0.57] & 0.46 [0.25,  1.17]\\
    \bottomrule
    \end{tabular}}
    \label{tab:mediation_indirect_effects}
    

\end{table}

Table~\ref{tab:mediation_indirect_effects} shows that when the perceptions of risks and benefits both had significant indirect effects, the effect sizes of the two benefit factors were almost always larger than the security and privacy risk factor.
For \textit{Age} and \textit{Transit use}, security risks even had a negative indirect effect, which means although these two independent factors had significant effects on security risk perceptions, the effects on other mediator variables were larger and had an opposite direction.
Therefore, we conclude that one's perceptions about the benefits of COVID-19 contact-tracing apps are more powerful determinants of app installation intentions than the perceptions about the security and privacy risks caused by the app.

Furthermore, we learned from Table~\ref{tab:mediation_indirect_effects} that \textit{Perceived adoption} often had an even larger effect size than the two benefit factors.
This result is not surprising as we already knew the efficacy of contact-tracing apps largely depends on whether it can achieve a widespread adoption.
If a person does not have enough confidence in having enough people installing a contact-tracing app, they may refrain from installing it themselves.

Figure~\ref{fig:sem} provides more information about two parts of an indirect effect: the correlation between the independent variable (the first-row nodes) and the mediator variables (the four second-row nodes on the right) and the correlation between the mediator variables and the outcome variable \textit{Install app} intention rating.
This could help us gain more understanding in the results of RQ1 and RQ2.
For example, we can see the negative correlation between being an essential worker and the installation intention rating could be partly attributed to the decreased perception of societal benefits.
However, we want to note that these four mediators were not able to explain all the effects.
For example, none of them had a significant indirect effect for explaining why Hispanics had significantly higher intentions to install the app than Whites, which requires further investigation by future work.

\section{Discussion}

Our research has several key practical implications on the design, marketing, and deployment of COVID-19 contact-tracing apps in the U.S., many of which could also apply in broader contexts such as strategies to increase adoption for digital technologies to help contain the spread of COVID-19 and building effective contact-tracing apps for infectious diseases in general.

\subsection{Design Contact-Tracing Apps to Match User Preferences}

Overall, our regression analysis showed that app design choices such as decentralized vs. centralized architecture, location use, who provides the app, and disclosures about app security risks had very small effects on participants' adoption intentions of COVID-19 contact-tracing apps (RQ1, Section~\ref{sec:RQ1_results}).
Since the baseline levels in our study represent the current design of contact-tracing apps in the U.S. (State-level, decentralized architecture, location collection not permitted), which features the strictest restrictions in data use, our results convey a positive signal that U.S. mobile users are open to or may even slightly prefer alternative designs that collect more sensitive data in a privacy-friendly way and offer additional benefits.
Participants also showed similar adoption intentions for app providers other than state health authorities, which suggests that using a piecemeal solution that leverages resources from different entities (e.g., Google/Apple OS-level support, apps provided by employers or schools) to complement the systematic yet slow responses from state-level authorities as proposed by~\citet{blasimme2020s} is a viable approach.

The few factors related to app design choices that had significant effects on adoption intentions also point out the sweet-spot in the current design space of contact-tracing apps to optimize for app adoption.
For the ``\textit{location uploaded}'' feature, although the current GAEN API does not allow collecting location directly in the same app, researchers have proposed creative solutions to gather information about places that infected users visited without logging location traces at an individual level~\cite{culler2020covista}.
The key idea is to treat places as people so the GAEN API could be extended to monitor a place's exposure to infected users and gather anonymized location traces of infected users at an aggregated level.
We consider this work a promising solution as it greatly reduces the security risk when maintaining the benefits that seem to be very attractive to users according to our results.

For the study around the security risk presentation, we learned that people were more concerned about the risk of secondary data use (which is more of an issue for centralized architectures), while less concerned about the risk of re-identification (one of the few security risks that decentralized apps are vulnerable to).
These results provide more empirical evidence to support the current deployment of decentralized architectures for contact-tracing apps.
Furthermore, as our results suggest that priming users about security risks does not reduce their app adoption intentions in most situations, app developers should be more candid about the possible security risks when presenting contact-tracing apps to users to help them make informed decisions.

\subsection{Consider Individual Differences in App Design and Marketing Strategies}


Contrary to the small effects of app design choices, we found individual differences had large effects on adoption intentions of COVID-19 contact-tracing apps.
First, we found people with higher prosocialness, higher COVID-19 risk perceptions and higher technology readiness are significantly more inclined to install and use contact-tracing apps.
This shows an marketing opportunity of contact-tracing apps to appeal to people with these characteristics by emphasizing related values such as helping the society combat the disease, helping protect yourself and other people, and taking advantage of the new technology to alleviate the work of human contact tracers.

Second, we found certain demographic groups had significantly higher or lower adoption intentions than other people regardless of the app design choices (RQ2.1-2.6, Section~\ref{sec:individual_difference_main_effect_results}).
Some findings show positive signals for the effectiveness of COVID-19 apps.
For example, public transit use is positively correlated with the intentions to install COVID-19 contact-tracing apps, which corresponds to one of the scenarios that these apps are expected to be most useful for.
Some of these findings are particularly concerning.
For example, older people had significantly lower intentions to install COVID-19 contact-tracing apps although they are at higher risk for severe illness from COVID-19.
Similarly, essential workers also had significantly lower intentions to install COVID-19 contact-tracing apps although they are at higher risk for exposure to COVID-19.
With our mediation analysis results (Section~\ref{sec:mediation_results}), we speculate that the lower installation intentions of older people could be because they were less tech-savvy and did not feel this technical solution provides much benefit to them.
For essential workers, the mediation analysis only showed a significant indirect effect through a reduction in the perceived societal benefit rating of the app.
We hope future research could conduct qualitative studies regarding the adoption intentions of essential workers in particular to provide better explanations about their preferences and rationales.

Third, we found different demographic groups had different preferences among two app design choices (RQ2.7, Section~\ref{sec:individual_difference_interaction_effect_results}).
Although these interaction effects did not change the general trends of adoption intentions for different demographic groups, we want to caution potential developers of contact-tracing apps of the unequal effects of certain app design choices on different demographic groups.
For example, although introducing location features sometimes increased the adoption intentions of participants in general, many essential workers and health workers seemed to prefer apps that do not collect location over those that do.
We speculate this may be because there is a greater privacy risk related to essential workers as their job require them to go outside and visit more places than other people.
This suggests that if app designers do want to incorporate location features for more public health benefits, the enabling of these features should be completely voluntary and require users to explicit opt in.
By protecting these vulnerable groups, we could also help better protect the general population due to the increase in adoption rate of people who are at higher risks of getting exposed.

For people living in rural areas, installation intention was drastically lower for apps developed by a large tech company than for apps developed by their state health authorities.
That is to say, contact-tracing apps developed by a large tech company may not be as effective in rural areas as in urban areas.
Note that in real world, the app provider may not be as obvious as in the app description of our study, which means that user's perceived app provider could have similar effect on their adoption intentions as the effects of app provider tested in our study.
Since current U.S. contact-tracing apps are all built with the GAEN API provided by Google and Apple, it is important for the marketing of the app to clearly convey to users who built the app and who has access to their data.

\subsection{Emphasize Public Health Benefits to Promote Contact-Tracing App Adoption}

The findings of our mediation analysis showed that although both security and privacy risks and public health benefits had significant indirect effects, the indirect effects of perceptions about contact-tracing apps' benefits (i.e., protecting the users themselves and the societal benefit of slowing the spread of COVID-19) were consistently larger than the indirect effects of perceived security and privacy risks.
This suggests that emphasizing the apps' benefits could increase user awareness of these benefits and drive more adoption, while efforts to decrease user awareness of security and privacy risks are likely to have less impact.
This result echos \citet{trang2020one}'s findings that the variations of app description in terms of benefits provided by the app had a larger effect size than variations in terms of privacy protection levels.

Accordingly, we derive two recommendations for designing and deploying COVID-19 contact-tracing apps.
First, contact-tracing app designers need to make sure the system works accurately, so that it actually offers key benefits. Opt-in features (e.g., progressive requests of location data) could allow users who are willing to contribute more data to obtain more useful features while enabling users who are more concerned about the security and privacy risks to share only the minimum amount of data.
Second, contact-tracing app design and marketing should also serve an educational purpose and emphasize more on the public health benefits both to the user themselves and to the society.
In addition to providing clear app descriptions, providing basic statistics using proper visualizations to help users get a better sense of how the app works in real life is also a direction worth exploring.

\subsection{Methodological Limitations}
This research has several limitations.
First, because our study tested hypothetical app designs to achieve a thorough and systematic exploration of the design space, we  could only investigate people's adoption intentions rather than their actual behaviors.
Therefore, our findings may not fully represent the corresponding actions people take for a real-world contact-tracing app.

Second, due to the constraint of the survey length, we had to trade off some details in app description and questions, such as to what extent the battery life is affected. Since the main goal of our study is to understand and compare the implications on adoption intentions of a wide range of factors, we consider this type of trade-off acceptable and would like to leave a more in-depth examination of specific factors for future work.

Third, users were reading app descriptions that presented more app design and implementation details (even including security risks in some conditions), which contained more information than they can obtain in real-world situations. This could affect the generalizability of the results. Although our findings suggest contact-tracing app providers should be more open about what benefits the app offers to motivate more adoption and what potential risks the app can cause to give people more transparency when not heavily discouraging their interests in using the app.

Fourth, We only surveyed mobile users and people aged over 18. So the findings may not generalize to people who are not using a mobile phone (but could use other approaches, such as IoT devices or infrastructure, to participate in digital contact tracing~\cite{Polenta_2020,tedeschi2020iotrace,hu2020iotbased}) and minors.
Also, we only surveyed U.S. people, which means the estimates of app adoption rate may not generalize to other contexts.

Lastly, due to the general limitations of quantitative study methodologies, we could not fully uncover the nuances in people's rationales behind their perceptions and adoption intentions, such as why Hispanic people and Black people had higher adoption intentions in some situations and why essential workers were less willing to install contact-tracing apps.
We hope future work could investigate these aspects specifically.

\section{Conclusion}
In this research, we conducted a national scale survey experiment ($N=\samplesize{}$) in the U.S.\ following a between-subjects factorial design to examine the effects of app design choices and individual differences on the adoption intentions of COVID-19 contact-tracing apps and how participants' perceptions of security and privacy risk, public health benefit, and community adoption rate mediate these effects.
Our results showed that individual differences had a larger impact on participants' app adoption intentions than app design choices, and both app design choices and individual differences affect the adoption intentions more through the perceptions of public health benefit and community adoption rate than perceptions of security and privacy risk.
Based on these findings, we derived practical implications on app design, marketing, and deployment.
Specifically, we identified sweetspots in the contact-tracing design space that could drive higher adoption.
We discussed app design considerations and marketing strategies with regards to individual differences, especially the importance of paying attention to protecting certain vulnerable groups such as essential workers, health workers, and people living in rural areas when designing and promoting the app.
Lastly, we emphasized public health benefit as an effective leverage to promote contact-tracing app adoption.


\section{Acknowledgement}
The authors would like to acknowledge Cori Faklaris, Ruotong Wang, and Laura Dabbish for their help on the study design.

This work is supported in part by CMU Block Center and the National Science Foundation under Grant No. 1801472.

\bibliographystyle{ACM-Reference-Format}
\bibliography{sample-authordraft}

\appendix

\section{App description example}

\begin{figure}
    \centering
    \includegraphics[width=0.9\linewidth]{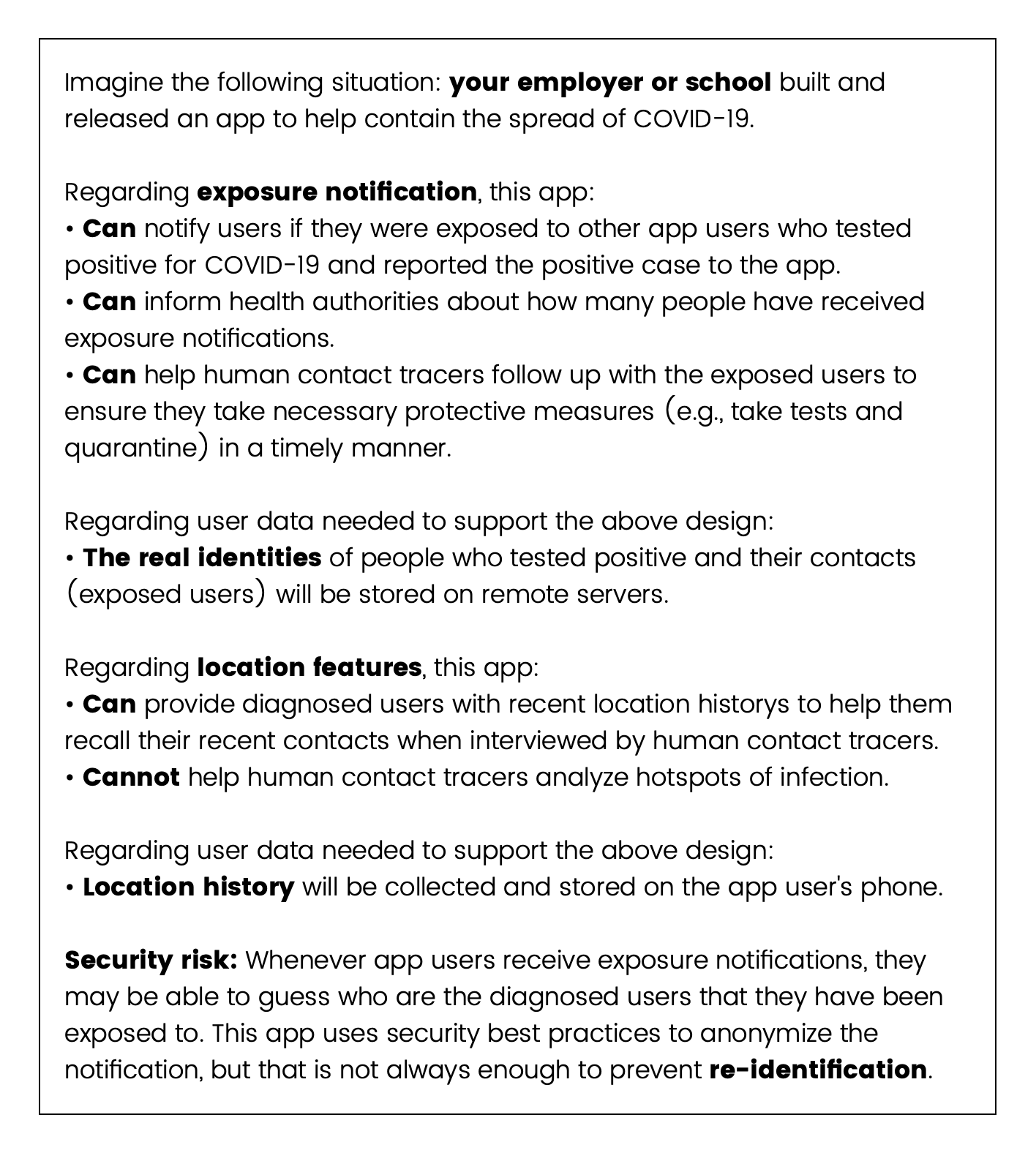}
    \caption{App description example.}
    \label{fig:example_app_description}
\end{figure}

\section{Additional statistical analysis results}
\begin{sidewaystable}[htbp]
    \caption{The complete results of our interactive regression model for the app installation intention ratings. We tested the interaction effects between app design choice factors and individual difference factors. To save space, we conflate the intercept values and main effect values into the first row and the first column.}
    \vspace{0.5em}
    \centering
    \resizebox{\linewidth}{!}{%
    \begin{threeparttable}
    \centering
    \begin{tabular}{p{0.25\linewidth} | p{0.13\linewidth}|
    p{0.13\linewidth}p{0.13\linewidth}p{0.13\linewidth} | 
    p{0.13\linewidth}p{0.13\linewidth} | 
    p{0.13\linewidth}p{0.13\linewidth} | 
    p{0.13\linewidth}p{0.13\linewidth}p{0.14\linewidth}}
    \toprule
    Individual & \multirow{2}{2cm}{Intercept/\newline Main Effect} & \multicolumn{3}{c|}{App Provider (State health authorities=0)} & \multicolumn{2}{c|}{Proximity (Decentralized=0)} & \multicolumn{2}{c|}{Location Use (None=0)} & \multicolumn{3}{c}{Security risk (None=0)}  \\ \cline{3-12}
                  difference &             & Federal health authorities & Tech company & Employer or school &    Anonymized Centralized   &  Identified Centralized  & Location on device & Location uploaded & Data breach risk & Secondary use risk & Re-identification risk \\ \midrule
    Intercept/Main effect & -1.066(0.833) & 1.172(0.668) & -0.352(0.659) & 0.128(0.651) & -0.989(0.562) & -1.791**(0.564) & -0.471(0.573) & 0.301(0.557) & -0.218(0.665) & 0.043(0.647) & 0.374(0.687)\\
    Age & -0.013(0.008) & -0.012(0.007) & 0.002(0.007) & 0.005(0.007) & 0.006(0.006) & 0.007(0.006) & -0.000(0.006) & -0.004(0.006) & 0.008(0.007) & -0.006(0.007) & -0.005(0.007)\\
    Gender (Male=0) \\
    \hspace{3mm} Female & -0.290(0.262) & -0.350(0.229) & -0.621**(0.231) & -0.200(0.233) & 0.029(0.203) & 0.624**(0.197) & 0.316(0.198) & 0.227(0.201) & 0.068(0.223) & -0.471*(0.221) & -0.239(0.236)\\
    Race (White=0) \\
    \hspace{3mm} Asian & 0.340(0.549) & 0.387(0.448) & -0.030(0.448) & 0.075(0.498) & -0.211(0.371) & -0.347(0.402) & -0.008(0.408) & 0.024(0.386) & 0.081(0.443) & -0.113(0.487) & -0.273(0.441)\\
    \hspace{3mm} Black/African American & -0.325(0.385) & 0.047(0.345) & -0.205(0.335) & -0.193(0.332) & 0.662*(0.297) & 0.057(0.292) & 0.141(0.301) & 0.142(0.296) & 0.147(0.324) & 0.603(0.329) & 0.178(0.348)\\
    \hspace{3mm} Hispanic/Latino & 0.245(0.388) & -0.365(0.351) & 0.092(0.349) & 0.392(0.355) & -0.155(0.310) & -0.410(0.296) & 0.315(0.302) & 0.411(0.312) & -0.151(0.335) & 0.124(0.353) & 0.013(0.371)\\
    Education & -0.023(0.077) & -0.054(0.066) & -0.022(0.066) & -0.056(0.066) & 0.085(0.058) & 0.118*(0.057) & 0.090(0.057) & 0.042(0.057) & -0.066(0.066) & 0.014(0.064) & -0.017(0.069)\\
    Household Income & 0.176*(0.077) & -0.077(0.068) & -0.111(0.067) & 0.019(0.069) & -0.036(0.058) & 0.027(0.059) & -0.043(0.059) & 0.020(0.058) & 0.020(0.066) & -0.055(0.066) & -0.002(0.070)\\
    Essential worker & 0.367(0.319) & -0.092(0.286) & 0.099(0.280) & 0.132(0.282) & -0.089(0.249) & -0.217(0.238) & -0.160(0.241) & -0.544*(0.246) & -0.126(0.287) & -0.427(0.276) & -0.484(0.290)\\
    Health worker & -0.662(0.529) & 0.493(0.447) & 0.116(0.433) & 0.007(0.442) & 0.753(0.391) & 0.319(0.377) & -0.939*(0.410) & -0.403(0.381) & 0.644(0.434) & 1.046*(0.437) & 0.878*(0.432)\\
    Public transit use & 0.052(0.126) & -0.125(0.107) & 0.042(0.108) & -0.007(0.106) & 0.055(0.091) & 0.123(0.093) & 0.061(0.091) & 0.104(0.093) & 0.031(0.104) & 0.063(0.106) & -0.047(0.106)\\
    Urban area percentage & 0.003(0.006) & 0.007(0.005) & 0.012*(0.005) & -0.000(0.005) & 0.000(0.004) & 0.002(0.004) & -0.001(0.004) & -0.005(0.004) & 0.000(0.005) & 0.003(0.005) & 0.001(0.005)\\
    Prosocialness & 0.404***(0.054)\\
    General privacy concern & 0.679***(0.049)\\
    COVID-19 risk perception & -0.100*(0.048)\\
    Technology readiness & 0.680***(0.068)\\
    \bottomrule
    \end{tabular}
    \begin{tablenotes}[para,flushright]
    Note: * p$<$0.05; ** p$<$0.01; *** p$<$0.001
    \end{tablenotes}
    \label{tab:interaction_regression_results}
    \end{threeparttable}}
\end{sidewaystable}

\end{document}